\documentclass[twocolumn,preprintnumbers,amsmath,amssymb]{revtex4-1}
\usepackage{epsfig}
\usepackage{amsmath}
\usepackage{color}
\begin{document}
\title{Homogeneous ice nucleation evaluated for several water models}
\author{ J. R. Espinosa, E. Sanz, C. Valeriani and C. Vega }
\affiliation{Departamento de Qu\'{\i}mica F\'{\i}sica,
Facultad de Ciencias Qu\'{\i}micas, Universidad Complutense de Madrid,
28040 Madrid, Spain}
\date{\today}
\begin{abstract}

In this work we evaluate by means of computer simulations 
 the rate for ice homogeneous nucleation for several water models such as TIP4P, TIP4P/2005,TIP4P/ICE and mW 
 (following the same procedure as in  J.A.C.S. {\bf 135} 15008 (2013)) 
 in a broad temperature range. 
We estimate the ice-liquid interfacial free-energy, and conclude that 
for all water models $\gamma$ decreases as the temperature decreases.
Extrapolating our results to the 
melting temperature, we obtain a value of the interfacial free-energy between 25 and 32 mN/m 
in reasonable agreement with the reported experimental values. Moreover,  
we observe that the values of   $\gamma$ depend  on the chosen 
water model and this is a key factor when numerically evaluating nucleation rates,  
given that the kinetic prefactor is quite similar for all water models with the exception of the mW 
(due to the absence of hydrogens). 
Somewhat surprisingly the estimates of the nucleation rates found in this work for TIP4P/2005 are slightly higher than 
those of the mW model, even though the former has explicit hydrogens. Our results 
suggest that it may be possible to observe spontaneous crystallization of TIP4P/2005 at about 60K below 
the melting point.

\end{abstract}

\maketitle


\section{Introduction}

When liquid water is super-cooled to below its melting point, it becomes metastable and eventually 
freezes into its thermodynamically stable phase (ice).
On the one hand, in the presence of impurities, this phase transition occurs quite easily (this is the reason why 
 ice will appear in your refrigerator only after few hours). 
On the other hand, 
in the absence of impurities, metastable liquid water can survive even at temperatures well below 
the melting point, until homogeneous nucleation takes place and water is transformed into ice.
Homogeneous nucleation is an activated process, given that  the system has to overcome 
a nucleation free-energy barrier and to form a critical ice cluster 
in order to crystallize \cite{kelton}.

By performing experiments with micrometer-size water droplets, it has been possible
to prepare metastable liquid water at temperatures down to 235K \cite{pruppacher1995,taborek,stockel2005,murray2010}. 
Below this temperature (known as the homogeneous nucleation temperature) water freezes in a few seconds. 
Such experiments  permits one  to experimentally determine the nucleation rate J (i.e the number of ice critical clusters per
unit of volume and time) for temperatures between 235K and 242K, with values of J  defined within less than three orders of 
magnitude. Outside this range, it has not been possible to experimentally determine the nucleation rate, either because 
 it is too large (below 235K) or  too small (above 242K). 
Given that J is known only in a narrow temperature range, to estimate its values outside such range  \cite{murray2010}
 classical nucleation theory (CNT)  \cite{kelton} could provide reasonable predictions, 
since the main  ingredients needed are the  interfacial free-energy of the liquid-ice interface at coexistence ($\gamma$) and the kinetic prefactor.
 However, on the one hand, even though  $\gamma$  could in principle be experimentally measured, 
  its reported values (so far) range from 25 to 35 mN/m \cite{pusztai2002}; on the other hand, the kinetic prefactor 
  is not known experimentally.
  


For these reasons, we believe that computer simulation could give a reasonable contribution in this context, since they could help 
both in determining the value of $\gamma$ and  evaluating 
the homogeneous nucleation rate  over a broader temperature range.
As far as we are aware, little work has been devoted numerically  to compute  $\gamma$ for the ice-water interface: 
the only exception being Ref.\cite{gammadavid,gammadavid_old}, where  $\gamma$ was calculated 
  at the melting point for several water models.

Moreover, work still needs to be done  to estimate  ice nucleation rates by means of numerical simulations.  
First of all, in order to know the amount of supercooling of liquid water (which determines the nucleation rate)
one needs to  know  the melting temperature. 
However, until  2005~\cite{JCP_2005_122_114507},  the melting point of most  water models had not been calculated. 
The first pioneering numerical paper on ice nucleation was that of  Matsumoto et al \cite{N_2002_416_00409}, where  
 spontaneous crystallization was observed at 230K for a system of 500 molecules at a pressure of about -1000 bar using the TIP4P model 
  \cite{JCP_1983_79_00926}.  Later on, for the same water model,  the nucleation free-energy barrier  had been calculated at 180K in Ref. 
\cite{PhysRevLett.90.158301},\cite{quigley:154518} and \cite{anwar_tip4p_nucleation}.  
The nucleation rate  has been also recently computed for the mW water model\cite{doi:10.1021/jp805227c}  by Li et al \cite{galli_mw} 
using forward flux sampling between 240K and 220K, by Reinhardt and Doye \cite{doye12}  using umbrella sampling at 220K. At 220K 
the value of J computed for mW by both groups differs by 5 orders of magnitude.
This difference is somewhat larger than the 
expected statistical uncertainty for nucleation rates (which is expected to be of 1-2 orders of magnitude). Although both groups used different rare-events techniques 
the origin of the discrepancy it is not clear as for other systems the values of J computed from 
forward flux sampling and umbrella sampling seems to be in better agreement\cite{hs_filion}.
For the mW model using brute force simulations at 208K Moore and Molinero \cite{nature_valeria}
were able to nucleate ice spontaneously in about 100ns in a system of 5000 molecules, 
leading to a rate of about $10^{32} m^{-3}s^{-1}$. 
 In 2013, our group estimated the value of J and $\gamma$ for other two water models, 
TIP4P/2005 \cite{TIP4P2005} and TIP4P/ICE \cite{TIP4PICE},   at low/moderate supercooling 
 using the "seeding technique" \cite{bai2,bai:124707}   together with CNT \cite{jacs2013}. 
Even though the main advantage of the seeding technique is that it allows one to estimate the nucleation free-energy barrier even at 
 moderate supercooling (differently from  more rigorous numerical techniques such as umbrella sampling or forward flux sampling 
that  might be CPU-time consuming at such temperatures), 
its main disadvantage is that it combines precise simulation results with an approximate theoretical formalism. 
The nucleation rates evaluated for both TIP4P/2005 and TIP4P/ICE water models 
 \cite{jacs2013} were in reasonable agreement with experiments. 
 However, this agreement may have been due to a fortuitous cancellation of errors, occurring when an approximate 
water model is used in combination with  an approximate technique. 
Therefore, in this work we will apply the same technique as in Ref.\cite{jacs2013}  
to estimate ice nucleation rate using other water models, such as 
 TIP4P  \cite{JCP_1983_79_00926} and mW \cite{doi:10.1021/jp805227c}.

In what follows, we will first provide more technical details about our previous work \cite{jacs2013}. 
Next, we will analyze the differences in the estimates of $\gamma$ for several water models, 
and observe that  values of $\gamma$ change significantly from a 
water model to another: even though for all water models $\gamma$ decreases as the temperature decreases. 
We then compute the kinetic prefactor, and conclude that it 
 is quite similar for all water models, with the exception of mW for which it
is about three orders of magnitude larger: this is certainly due to the lack of hydrogens in the model. 
However, being this difference significant, it is  $\gamma$ that plays the central role in 
determining the nucleation rates. 
To conclude,  
 we evaluate J and compare the results obtained for each water model. 
In particular we will focus on the mW model potential 
to determine whether the nucleation rate can be enhanced compared to other water  models.
We first observe that J estimated with the seeding technique compares nicely to the 
 values of J reported for the same mW model in the literature  (to within 5-6 orders of magnitude which is the expected uncertainty at high supercooling). 
Somewhat surprisingly, estimates of the nucleation rates  for TIP4P/2005 are slightly higher than 
those for the mW model (even though the former has explicit hydrogens). 
The results of this work suggest that it may be possible to observe spontaneous crystallization of TIP4P/2005 at about 57K below 
the melting point (i.e. 195K). Given that nucleation rate at 230K is very small, nucleation is  
not likely to be observed in computer simulations for TIP4P/2005.
  At such low temperature (and room pressure) a maximum in the compressibility has been found for this model by Abascal and Vega \cite{abascal10,abascal11} and Bresme et al. \cite{bresme2014} thus providing a point of the Widom line. 
 The results of this work support the existence of the Widom line 
  for TIP4P/2005, and that this line is not due to the  transient formation of ice\cite{limmer_2}.


\section{Methodology}

\subsection{The "seeding" technique}

The technique first proposed by Bai et al. \cite{bai2,bai:124707}  
consisted of inserting a solid cluster in a supercooled Lennard-Jones fluid, determining the temperature at which the cluster was critical (i.e 
where it can freeze or melt with equal probability). 
We shall denote this technique as "seeding" , as it can be regarded as the
insertion of a seed of the stable phase (i.e the solid)  in the supercooled liquid. 

By assuming that classical nucleation theory can be used to describe and interpret the results obtained  for the 
critical cluster size, then the technique allows one to estimate of the interfacial free energy 
 $\gamma$ at the given thermodynamic conditions. According to CNT the critical cluster size $N_c$ is  
 \begin{equation}
\label{ncrit}
N_{c} = \frac{32 \pi \gamma^3}{3 \rho^2_s |\Delta \mu|^3}
\end{equation}
where $\rho_s$ is the number density of the solid phase (i.e ice Ih), 
 $\Delta \mu$ the chemical potential difference between
the solid and the fluid phase at the temperature at which the cluster is critical.  

Once the value of $\gamma$ has been determined via Eq. \ref{ncrit} one can estimate (once again using CNT) the free energy barrier for
nucleation from the expression :
\begin{equation}
\label{eq_G_cnt}
\Delta G_c  = \frac{16 \pi \gamma^3}{3 \rho^2_s |\Delta \mu|^2}.
\end{equation}
Finally, one can estimate nucleation rates. Following the approach described in detail by Auer and Frenkel \cite{auerjcp,kelton,Nature_2001_409_1020}, 
J can be obtained from the expression :
\begin{equation}
J=  \rho_f Z  f^{+} \exp(-\Delta G_{c}/(k_B T))
\label{eqrate}
\end{equation}
where ($\rho_f Z  f^{+}$) is the kinetic prefactor, with  
 $f^{+} $ the attachment rate of particles to the critical cluster, $\rho_f$ the number density of the fluid and $Z$ the
Zeldovich factor \cite{kelton}.  The CNT form of the Zeldovich factor is 
\begin{equation}
 Z=\sqrt{(|\Delta G^{''}|_{N_c}/(2\pi k_B T))} = \sqrt{|\Delta \mu|/(6 \pi k_B T N_c)} 
\label{zeldovich} 
\end{equation}
so that Z can be easily computed, once the size of the critical cluster $N_c$, the temperature at which 
it is critical  $T$ and the chemical potential difference  between the solid and the liquid are known. 
According to Ref. \cite{auerjcp,Nature_2001_409_1020,nacl_valeriani}, $f^{+}$ can be computed 
as a diffusion coefficient of the 
cluster size at the top of the barrier ( at the temperature at  which the cluster is critical): 
\begin{equation}
f^+=\frac{<(N(t)-N_c)^2>}{2t}. 
\label{eqattach}
\end{equation}
The seeding technique can be particularly useful at moderate supercooling, where estimating  the critical cluster size, 
the free-energy barrier height and the rate by more rigorous numerical techniques 
 would be very CPU-time consuming.

A similar approach has been recently  used by Pereyra et al 
\cite{carignano}, where the authors  determined the temperature 
at which a cylindrical ice slab would melt or grow, in Ref.
 \cite{knott12}, by Knott et al determined the critical cluster size in a nucleation study
of methane hydrate,  and in Ref. \cite{jacs2013}, where Sanz et al studied ice nucleation 
from supercooled water.

\subsubsection*{Drawbacks in the estimate of $\gamma$}

Admittedly, the way presented in Eq.(\ref{ncrit}) to estimate $\gamma$ is quite approximate, since 
it assumes that CNT is correct. 
The justification of this approach can be provided only a posteriori by comparison with more rigorous calculations. 

First of all, different crystal planes will have different values of $\gamma$, whereas   $\gamma$ computed according to 
 Eq.(\ref{ncrit}) does not take into account the different crystal planes and only corresponds to an average 
among them. 
On the one hand, it has been shown for several systems such as hard spheres, Lennard Jones and water 
\cite{gammadavid,gammadavid_old,anwar_tip4p_nucleation,JCP_2006_125_094710,davidchack:234701,davidchack:7651}
that comparing $\gamma$ computed for different planes results in differences 
 smaller than about five per cent. 
On the other hand, one may assume that the spherical interface  will represent the average value of $\gamma$ over different planes. 
In any case the relation between the value of $\gamma$ of a spherical cluster
with that computed for a planar interface  is not completely clear. 

To conclude, our calculations of $\gamma$ rely on the assumption that the shape of the 
cluster is spherical. Visual inspection of our molecular dynamics trajectories suggests
that this is indeed a reasonable approximation. 

\subsection{Distinguishing between liquid and ice-like molecules}

As in our previous work\cite{jacs2013}, in order to identify molecules as liquid or ice Ih-like,  
we have used the $\bar{q}_{6}$ order parameter proposed in Ref. \cite{dellago}: 
molecules with $\bar{q}_{6}$ larger than 0.358 will be classified as solid (ice Ih) and those with smaller 
values of as liquid-like.
\begin{figure}[h!]
\includegraphics[width=0.4\textwidth,clip=]{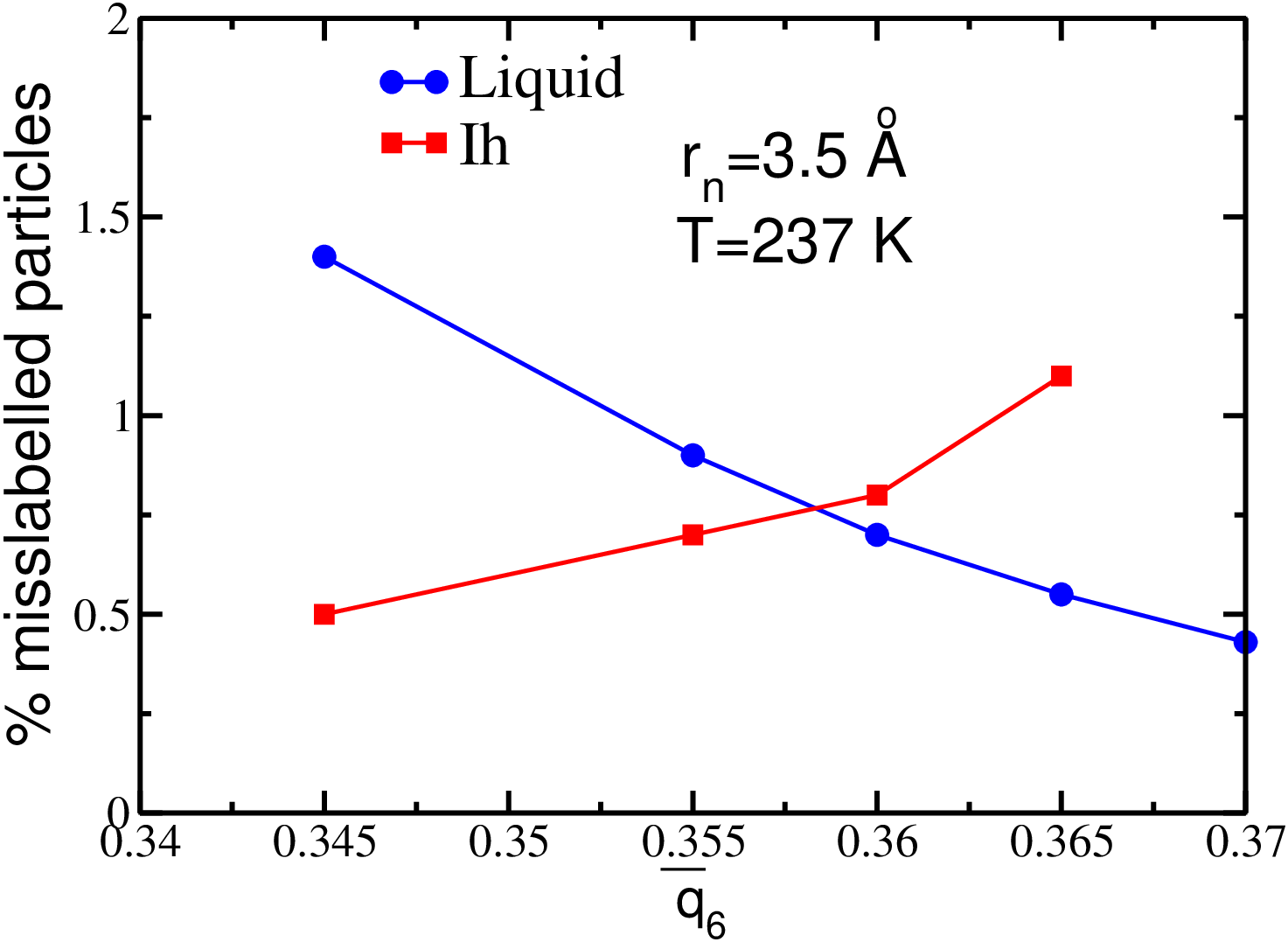} 
\caption{Percentage of mislabelled particles according to  $\bar{q}_{6}$, 
evaluated for bulk ice Ih and bulk liquid water 
at 237K and 1bar (for TIP4P/2005), using the first minimum of the $g(r)$ 
as a cutoff for the calculation of $\bar{q}_{6}$. For ice Ih mislabelled particles are those with a value
of the order parameter smaller than $\bar{q}_{6}$. For liquid water mislabelled particles are those with 
a value of the order parameter larger than $\bar{q}_{6}$.   
}
\label{q6q4}
\end{figure}

Following this criterion, we conclude that only about $0.7$  per cent of bulk ice Ih molecules are wrongly 
identified as liquid-like, and vice versa $0.7$ per cent of bulk liquid molecules identified as solid-like (see Fig. \ref{q6q4} ) .
Since ice Ih and supercooled water have a quite similar structure,  one may neglect this small mislabelling 
(furthermore it is very difficult to find  order parameters with smaller mislabelling). 

\subsubsection*{Drawbacks in the estimate of $N_c$}
 
The solid-fluid interface at the nucleus is not sharp, and we 
implicitly assume that the 
width of the interfacial region is very small relative to the size of the nucleus. This is of course, 
an approximation.  Order parameters are very useful to distinguish between bulk ice and bulk liquid, 
but it is by far more difficult to distinguish between liquid and solid molecules in the interfacial region.\cite{hs_filion,doye12,galli_mw} This constitutes a systematic source of error. 
 
Therefore, determining $N_c$ entails an uncertainty due to the interfacial molecules. 
Of course, the larger the clusters the smaller the amount of uncertainty in $N_c$, 
since the ratio of the number of molecules at the interface to those in the cluster's core decreases 
with the system size.  Whether the approach used in this work is reasonable  or not can only be tested at posteriori, 
by comparing the results of this work with those found in the literature.

\subsection{Our setup for the seeding technique}

In this work, by means of the seeding technique, we determine the 
temperature at which three clusters of different sizes are critical.  Three 
initial systems were obtained by inserting  spherical
ice Ih clusters of different sizes  in supercooled water (molecules overlapping with the cluster were removed). 
After inserting the cluster, we equilibrated its interface for about 0.2ns at 200 K, enough 
to equilibrate the interface but not to observe melting or growing of the cluster 
(which typically requires 2-20ns).  
After this 0.2ns the sizes of the ice cluster were  of about 7930, 3170 and 600 respectively for the three clusters
sizes considered in this work. 
In each system, the total number of water molecules  was about 20 times larger than the inserted 
cluster to avoid interactions between the cluster and its periodic images. 
Thus, the total number of molecules of water (considering both the ice Ih cluster and the  molecules
of the supercooled liquid ) were 182585, 76781 and 22712 respectively. 
In order to be able to simulate such rather large systems, we had to recur to supercomputer facilities.

Once the cluster is equilibrated, we performed MD runs at 
different temperatures and monitor the cluster size  to determine 
the temperature at which each cluster is critical.

 There is an additional point worthy of  comment concerning our initial setup. 
When implementing the seeding technique we use a starting cluster with Ih crystal
structure. Yet in recent work, both experimental  and numerical, have been strongly suggested that
initial ice nuclei contain stacking faults. This has resulted in recent
papers referring to stacking disordered ice I \cite{carignano_stacking,chinos_ice_growth,valeria_pccp_2011,kuhs_pnas_2012,malkin_pnas_2012,benet_2014}.
One may wonder about the consequences of this 
on the present study as it could have some impact on some relevant  quantities such as the
chemical potential difference, the interfacial free energy, and kinetic factors. 
This is an interesting point that deserves an independent study on its own.  However there is some 
indication that the impact of the presence of stacking faults in ice I on the final results may be rather small. 
Free energy calculations ( obtained from the Einstein crystal calculations) for ices Ih and Ic using the TIP4P/2005 model indicate 
that the free energy difference between these two solid phases is quite small\cite{conde_vega}. In addition, preliminary calculations
similar to those performed in this work, but inserting a cluster of pure Ic, reveal little differences with those obtained 
using a cluster of ice Ih\cite{alberto} ( suggesting that both the interfacial energy and the kinetic factors are quite similar 
for ices Ih and Ic).


\subsection{The chosen water model potentials}

In Ref.\cite{jacs2013}, we have studied both TIP4P/ICE and TIP4P/2005 water models, where  
MD runs were performed with Gromacs \cite{JMM_2001_07_0306} using a velocity-rescaling \cite{bussi07} thermostat and an istropic Parrinello-Rahman barostat \cite{JAP_1981_52_007182} 
with a relaxation time of about 2ps. 
The LJ term of the potential was truncated at 9 \AA\ and long range corrections 
were added to account for the truncation of the LJ part. Ewald sums ( with the PME technique \cite{CPL_2002_366_0537}) were used to deal with 
the electrostatic interactions. The real part of the electrostatic potential was also truncated at 9 \AA. 
In this work we shall extend
our previous study to the TIP4P model \cite{JCP_1983_79_00926}. The details of the simulations are similar to those
used in our previous work. In addition we have also performed simulations for the mW model of water \cite{mw}.  
Simulations for the mW model were performed  using the LAMMPS package \cite{lammps_program} 
In the mW water model hydrogens are not present, and tetrahedral ordering  is induced by using three body forces. 
The model has no charges, and due to the short range of the two and three body forces
it is computationally very fast. 

The comparison between the results of  TIP4P family models is of interest, as these models present the same charge
distribution (with one LJ center on the oxygen, two positive charges on each  H and a negative charge
on the H-O-H bisector) but differ in the strength of the hydrogen bond (increasing as TIP4P, TIP4P/2005 
and TIP4P/ICE) and thus in their melting points (increasing in the same order).

The mW model has recently become quite popular in nucleation studies (either  
 brute force \cite{nature_valeria} or using umbrella sampling or forward flux sampling techniques \cite{galli_mw,doye12}.
Therefore, we will use this model to test the validity of the seeding technique  and to 
 analyze whether the absence of hydrogens speeds up  the nucleation rate compared to 
other models where hydrogen  atoms are explicit.

 Let us finish this section with a final comment. 
 In this work we are using classical statistical mechanics ( i.e standard molecular dynamics simulations). 
 Since nucleation of ice occurs at low temperatures, where nuclear
  quantum effects gain importance, one may wonder about possible impact of such effects on nucleation studies of water. 
  The parameters of empirical potentials are typically obtained by forcing the model to reproduce 
  experimental properties within the framework of classical statistical mechanics. Thus the parameters 
  of empirical potentials  incorporate to some extend nuclear quantum effects in an effective way.
  That may explain the success of models like TIP4P/2005 to describe interfacial free energies and dynamic properties
  of real water.  As  will be shown in this work this strategy seems 
  to also be  successful when estimating  nucleation rates of water. 
  However, the properties of deuterated water (melting point, temperature of the maximum in density) 
  differ significantly from those of water indicating  that nuclear quantum effects are 
  important and this effect can not be captured by classical statistical mechanics (i.e within this framework the melting
  point does not depend on the mass associated with  the hydrogen atom). To capture isotopic effects in  nucleation 
  studies of water, it is necessary to have an accurate potential energy surface of water (obtained from 
  accurate electronic structure calculations) , and to incorporate 
  nuclear quantum effects. However we have shown recently that by using a modified 
  version of TIP4P/2005 (TIP4PQ/2005) in combination with path integral simulations, it is to describe reasonably well isotopic 
  effects in water\cite{mcbride09,vega10,mcbride12a,mcbride12b}. It would be interesting in the future to 
  pursue a study similar to that performed in this work, where TIP4PQ/2005 is used in combination with path 
  integral calculations to analyze isotopic effects on the nucleation of ice ( although this calculations would be 
  at least one order of magnitude more expensive than those performed in this work). 
  In any case the results of this work seem to indicate, that TIP4P/2005, in combination with classical simulations,
  seems to be reasonably successful in describing experimental values of the nucleation rates. Therefore the strategy
  of incorporating nuclear quantum effects via effective potentials does not seem too bad for this problem.

\section{Results}

Before presenting our main results, we summarize a few properties at the melting point 
of the chosen water potentials  (Table \ref{melting_properties}). 
\begin{table}[h!]
\caption{Melting temperature, ice Ih density \cite{vega09,vega11}, melting enthalpy and $\gamma$ at coexistence 
 (extrapolated from the results for the finite size 
clusters) for  TIP4P, TIP4P/2005, TIP4P/ICE, mW and experiments.}
\label{melting_properties} 
\centerline{
\begin{tabular}{ccccc}
\hline 
 model    &   $T_m$/K   &     $\rho_s$/$(gcm^{-3})$  &   $\Delta H_m$/(kcal/mol)  & $\gamma$/(mN/m)   \\
\hline
TIP4P & 230 & 0.94 & 1.05 & 25.6  \\
TIP4P/ICE & 272 & 0.906 & 1.29 & 30.8  \\
TIP4P/2005 & 252 & 0.921 & 1.16 & 29.0  \\
mW  & 274.6 & 0.978 & 1.26 & 29.6  \\
Experiment & 273.15 & 0.917 & 1.44 & 29 \\
\hline 
\end{tabular}}
\end{table}

All chosen water models differ in their properties at the melting point. No 
 water model is able to simultaneously reproduce the coexistence density, the melting temperature and
the melting enthalpy 
(even though TIP4P/ICE nicely reproduces 
the melting temperature and the solid density, it underestimates the melting enthalpy by about ten per cent). 
  The experimental density of ice Ih at the melting point is $0.92g/cm^3$ \cite{feistel06}. It is clear from the results of Table
 \ref{melting_properties} that the density of ice Ih is very well reproduced by TIP4P/2005 and TIP4P/ICE, and reasonably well by 
 TIP4P, whereas for mW the density of ice Ih is too high. 
Given that mW reproduces reasonably well the density of water at the melting point (i.e  1 $g/cm^3$) it turns out that 
 for this water model the density change from ice Ih to liquid water is only of about 2 per cent, considerably 
 smaller than that found in experiments where the density change is about 10 per cent. In other 
words, for mW , freezing is a weakly first order phase transition. 

Our main results for all water models are summarized in  Table \ref{jacs_table}. The runs used to determine the temperature at which each of the studied clusters becomes critical
are provided as Supplementary Material \cite{material_suplementario}.

\begin{table*}[t]
\caption{
Reported for a given cluster size  and water model are the corresponding supercooling, $\Delta T$/K, ice-Ih density, $\rho_s$/(g/cm$^3$), 
chemical potential difference between the liquid and the solid, $\Delta \mu (kcal/mol) $, number of particles in the cluster, $N_c$, attachment rate, 
$f^+$/s$^{-1}$, Zeldovich factor, Z, diffusion coefficient, D/(m${^2}$/s), $\lambda$/\AA, interfacial free energy, $\gamma$/(mN/m), height of the
nucleation free energy barrier, $\Delta G_c$/(k$_B$T), and decimal logarithm of the nucleation rate, $log_{10}$(J/(m$^{-3}$s$^{-1}$)).
Statistical errors for $\Delta G_c$ and $log_{10}(J)$  are shown in parenthesis. The uncertainty in $\Delta T$ is of about 2.5K, so that the errors 
in $\Delta \mu$, $\gamma$ and $\Delta G_c$ are of about 7 per cent. 
As discussed in the main text, if systematic errors are included, the error in $\gamma$ does
not increase much, but the error in  $\Delta G_c$ and  $log_{10} J$ presented in this Table should be multiplied by two. 
For the medium clusters  we have also included (in parenthesis) the value of the attachment rate and  $\lambda$ obtained 
using only times larger than 1.5 ns in the determination of the attachment rate.} 

\label{jacs_table} 
\centerline{
\begin{tabular}{cccccccccccc}
\hline
Model &  $\Delta T$ & $\rho_s$ & $\Delta \mu$ & $N_c$ & $f^+$ & Z & D & $\lambda$ & $\gamma$ & $\Delta G_c$ & $log_{10} J$ \\ 
\hline
Tip4p/ICE  & 14.5 & 0.908 & 0.0629 & 7926 & 6.9 $\cdot 10^{12}$  & 9.07 $\cdot 10^{-4}$ & 1.80 $\cdot 10^{-10}$ & 5.0 & 26.3 & 487(34) & -173(16)\\
Tip4p/ICE  & 19.5 & 0.909 & 0.0826 & 3167 & 2.9(2.6) $\cdot 10^{12}$  & 1.66 $\cdot 10^{-3}$ & 9.63 $\cdot 10^{-11}$ & 4.1(4.4) & 25.4 & 261(18) & -75(9)  \\
Tip4p/ICE  & 34.5 & 0.911 & 0.1335 &  600 & 3.0 $\cdot 10^{11}$  & 5.00 $\cdot 10^{-3}$ & 1.10 $\cdot 10^{-11}$ & 2.2 & 23.7 &  85(6) & 1(4)\\
\hline
Tip4p/2005  & 14.5 & 0.923 & 0.0612 & 7931 & 1.9 $\cdot 10^{12}$  & 9.31 $\cdot 10^{-4}$ & 1.48 $\cdot 10^{-10}$ & 6.4 & 25.9 & 515(36) & -186(17) \\
Tip4p/2005  & 19.5 & 0.924 & 0.0801 & 3170 & 1.2(1.3) $\cdot 10^{12}$  & 1.70 $\cdot 10^{-3}$ & 9.69 $\cdot 10^{-11}$ & 6.4 & 25.0  & 275(19) & -83(9)  \\
Tip4p/2005  & 29.5 & 0.925 & 0.1137 &  600 & 1.8 $\cdot 10^{11}$  & 4.76 $\cdot 10^{-3}$ & 3.31 $\cdot 10^{-11}$ & 6.5 & 20.4 &  77(5) &  3(3)   \\
\hline
Tip4p &  12.5 & 0.942 & 0.0515 & 7931 & 3.4 $\cdot 10^{13}$  & 8.92 $\cdot 10^{-4}$ & 1.44 $\cdot 10^{-10}$ & 2.0 & 22.0 & 472(33) & -166(15) \\
Tip4p &  17.5 & 0.943 & 0.0696 & 3170 & 4.0(5.6) $\cdot 10^{12}$  & 1.66 $\cdot 10^{-3}$ & 4.90 $\cdot 10^{-11}$ & 2.5 & 21.9 & 261(18) & -75(9)  \\
Tip4p &  27.5 & 0.944 & 0.1018 &  600 & 1.8 $\cdot 10^{11}$  & 4.73 $\cdot 10^{-3}$ & 1.06 $\cdot 10^{-11}$ & 3.1 & 18.5 &  76(5) &  4(3)   \\
\hline
\hline
mW &  14.6 & 0.980 & 0.0669 & 7926 & 9.0 $\cdot 10^{14}$  & 9.32 $\cdot 10^{-4}$ & 4.50 $\cdot 10^{-9}$  & 2.2 & 29.5 & 514(36) & -183(17) \\
mW &  19.6 & 0.981 & 0.0895 & 3167 & 2.3(2.5) $\cdot 10^{14}$  & 1.72 $\cdot 10^{-3}$ & 2.33 $\cdot 10^{-9}$  & 2.7 & 29.0 & 280(20) & -81(9)  \\
mW &  34.6 & 0.983 & 0.1553 &  600 & 1.1 $\cdot 10^{14}$  & 5.36 $\cdot 10^{-3}$ & 2.69 $\cdot 10^{-9}$  & 2.0 & 28.9 &  98(7) & -2(4)   \\
\hline
\end{tabular}}
\end{table*}

 \subsection{Ice Ih density}
 
As shown in Table \ref{jacs_table}, 
 the density of ice Ih increases as the temperature decreases and this is also  found in experiments ( at least up to 125K). 
  Below this temperature the experimental density of ice Ih is approximately constant. 
This is  a consequence of the third law of thermodynamics which implies that certain quantities such as the heat capacity or 
 the the coefficient of thermal expansion tends to zero
when the temperature goes to zero. Since the coefficient of thermal expansion goes to zero at low temperatures 
 that implies that the density of solid phases do not change much with temperature 
  at low temperatures (at constant pressure). These effects can not be 
 reproduced by classical simulations since their description would require the incorporation of nuclear
 quantum effects \cite{mcbride09,ramirez14}).

 \subsection{The chemical potential difference between the fluid and the solid, $\Delta \mu$}
 
 In order to determine the chemical potential difference between the liquid and the solid, we 
  perform NpT simulations  below melting for bulk ice Ih and liquid water. 
 Next, we compute the enthalpy in both systems and  perform thermodynamic integration to determine $\Delta \mu$ 
 (at coexistence, the chemical potential of the solid and liquid are the same).  
 \begin{figure}[h!]
\includegraphics[width=0.4\textwidth,clip=]{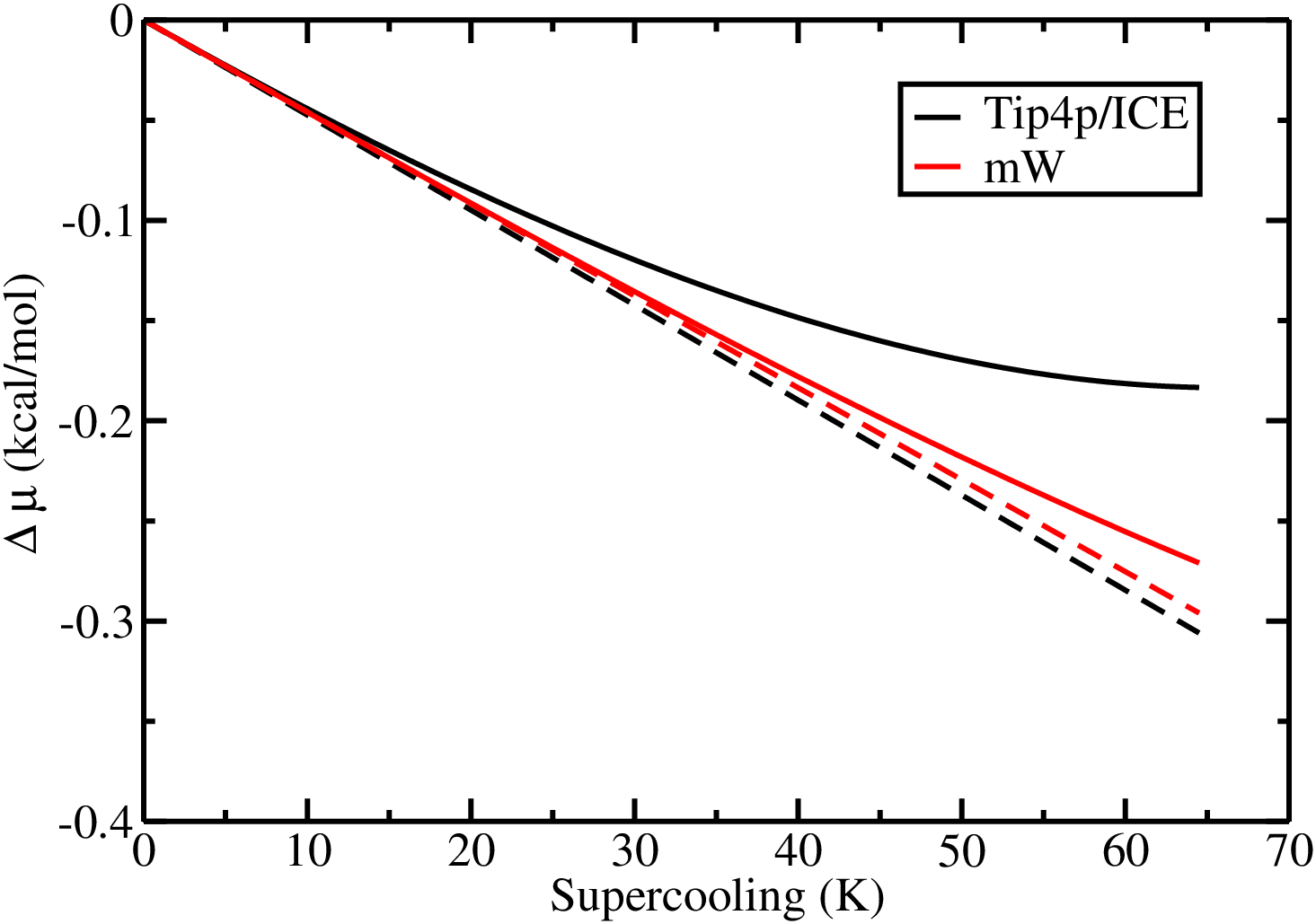} a)\\
\includegraphics[width=0.4\textwidth,clip=]{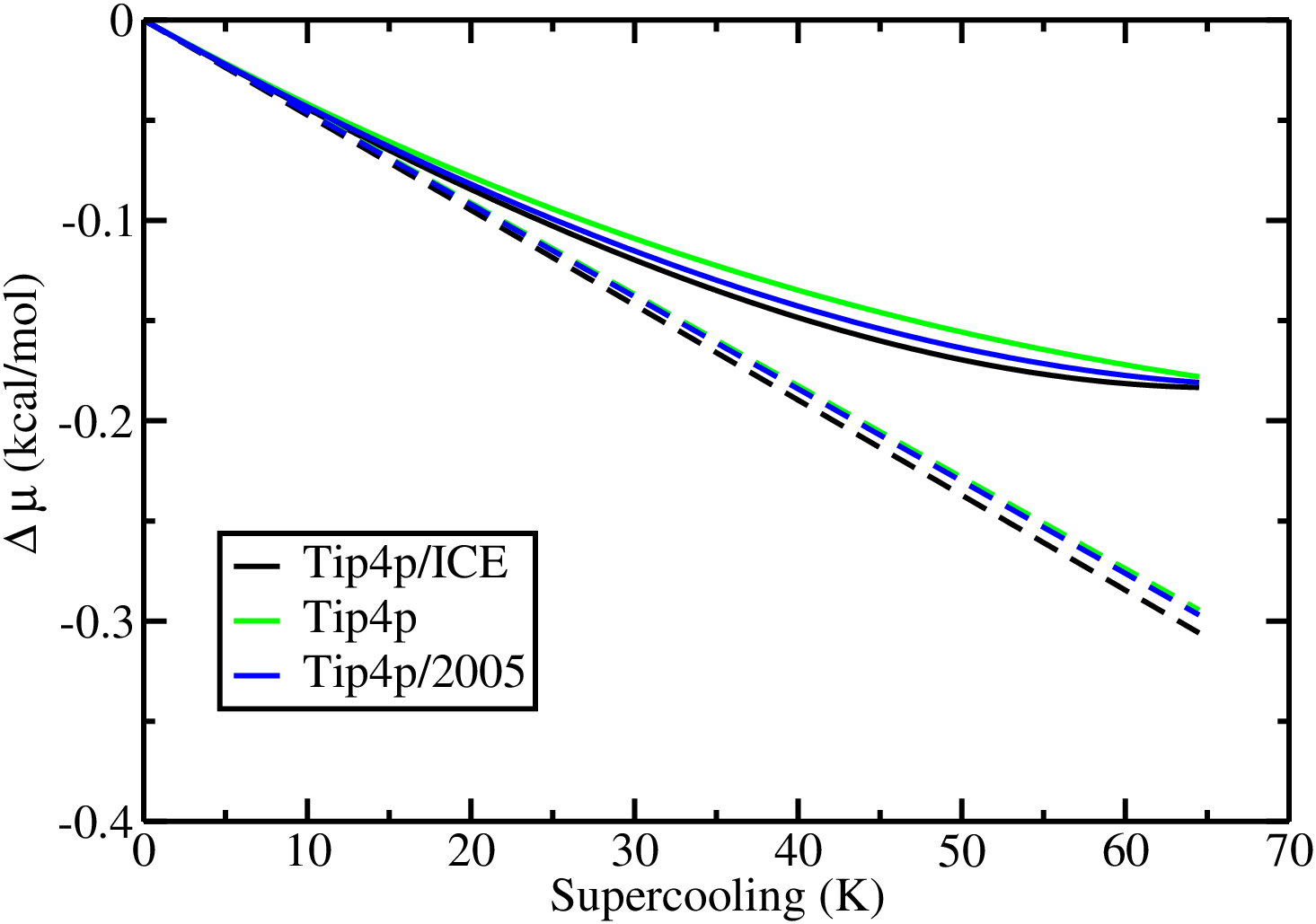} b) \\
\caption{$\Delta \mu$ obtained from thermodynamic integration (solid lines) and from Eq. \ref{deltamu_aprox} (dashed lines)
as a function of the supercooling $\Delta T$. 
a) Results for mW and TIP4P/ICE.  
b) Results for TIP4P, TIP4P/2005 and TIP4P/ICE. 
}
\label{delta_mu}
\end{figure}

As shown in Fig.\ref{delta_mu},   the value of $\Delta \mu$ is quite different for different models. 
$\Delta \mu$ can often be approximated using  the enthalpy change at melting  \cite{kelton}: 
\begin{equation}
\label{deltamu_aprox}
\Delta \mu   =  \Delta H_{m}    \left( 1 - \frac{T} {T_{m} } \right) 
\end{equation}
 As shown in Table \ref{melting_properties}, the enthalpy change at melting depends on the chosen model and   
 at the same supercooling   $\Delta \mu$  increases when using 
 TIP4P, TIP4P/2005,  TIP4P/ICE and  mW, respectively. 
  Even though Eq. \ref{deltamu_aprox} allows one to explain the results of Fig. \ref{delta_mu} for low supercoolings (where it 
 becomes basically exact), it cannot be safely used  for large supercoolings where 
  the value of $\Delta \mu$ rigorously  obtained from thermodynamic integration 
 visibly differs from that obtained via Eq. \ref{deltamu_aprox}. 
  The reason for this difference is that the enthalpy of  liquid water changes dramatically when water is supercooled, 
 as shown by the  increase in the heat capacity (which reaches a maximum 
 at the so called Widom line \cite{Kumar05062007}). 
  The only water model where the approximation works is the mW.
 For this model, the maximum in density is located at 250K, and the maximum  
 in the heat capacity is displaced to  lower temperatures. 
 To conclude, the value of $\Delta \mu$ at large supercooling 
 is sensitive to the thermodynamic behavior of supercooled water, and in particular to the location of the 
 maximum in the heat capacity (if any) with respect to the melting temperature.  

 \subsection{Determining   $N_c$}

To illustrate how the temperature at which the cluster is critical is determined we shall present 
one example for the mW model. 
\begin{figure}[h!]
\includegraphics[width=0.4\textwidth,clip=]{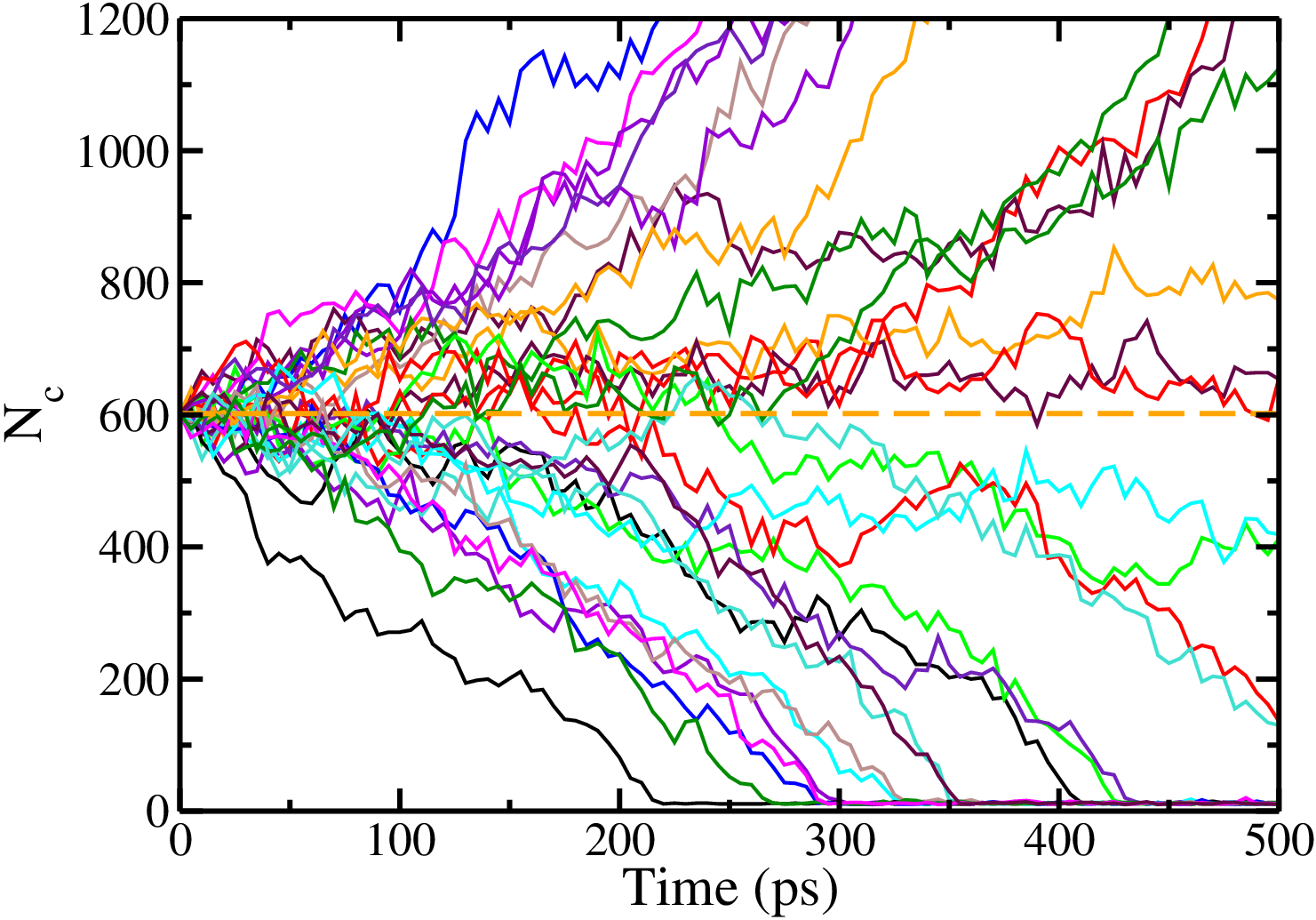} \\
\caption{ Time evolution of the cluster size  for the mW at T=240K and 1bar.  
                The size of the initial cluster was about 600 molecules. Results obtained
                for 30 independent trajectories are shown. }
\label{evolution_time_largest_cluster}
\end{figure}
In Fig.\ref{evolution_time_largest_cluster}, the time evolution  of the cluster containing 
600 ice molecules is shown for the mW model, at 1 bar and T=240K. At this temperature 
 the cluster is critical and in approximately half of the trajectories it melts, whereas in the other half it grows. 

 \subsection{The interfacial free energy, $\gamma$ }
 
\begin{figure}[h!]
\includegraphics[width=0.4\textwidth,clip=]{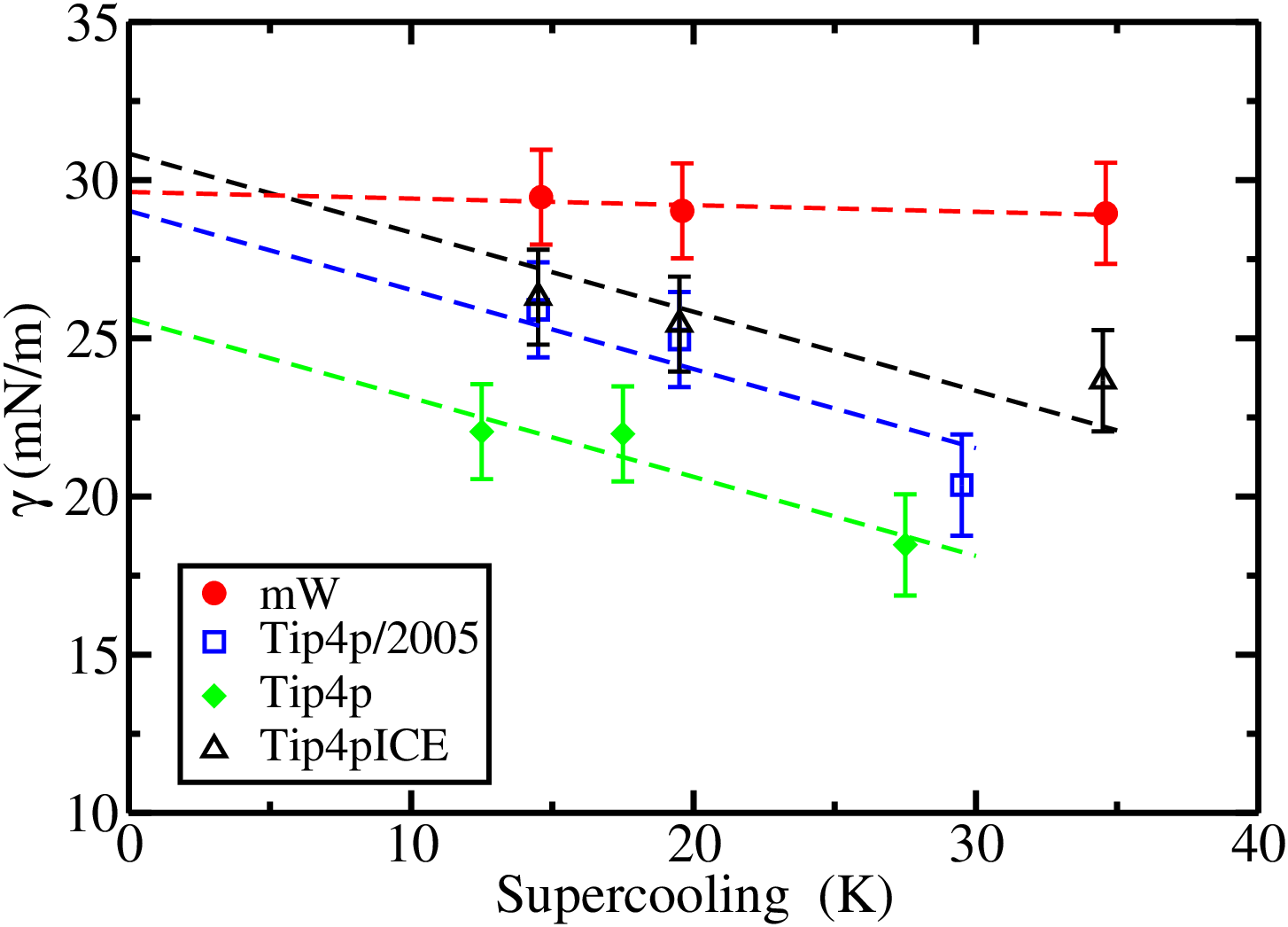}\\
\caption{Interfacial free energy between ice Ih and liquid water as a function of the degree of 
supercooling as obtained from the seeding technique in combination with CNT for TIP4P, TIP4P/2005, TIP4P/ICE and
mW. }
\label{sigma_vs_T}
\end{figure}

 By means of Eq.\ref{ncrit}, we have estimated $\gamma$  for each cluster size. 
In Fig.\ref{sigma_vs_T} the value of $\gamma$ is plotted as a function of the supercooling. 
    As can be seen for all models $\gamma$ decreases with the temperature (i.e decreases as one
  increases the supercooling).
   Our results are compatible with a possible linear decrease of $\gamma$ with T although a faster (than linear) decreases
  of $\gamma$ with T cannot be discarded.  
   The derivate of $\gamma$ with $\Delta T$ is the surface excess entropy.  
  We obtained a slope of 
  -0.13, -0.25 and -0.38 mN/(K.m) for TIP4P/ICE, TIP4P and TIP4P/2005 models. 
  These slopes have large error bars arising 
  from our uncertainty in the determination of $\gamma$. 
  To reduce such error bars we use the fact that all TIP4P-like models seem 
 to display similar behavior, 
  so we shall adopt the average
  slope, namely -0.25 mN/(K.m), for the three models.
  Such slope is in good agreement with the slope calculated in  Ref. \cite{doyejcp2013} 
  for the TIP4P/2005  ($-0.18$  mN/(K.m)).
  Experimentally there is no consensus neither on the value of $\gamma$ for the planar interface nor on the 
  change of $\gamma$ with the degree of supercooling (see fig.10 in the paper of Pruppacher \cite{pruppacher1995}). 
  In any case the slope reported here,  namely $-0.25 mN/(K.m)$, is roughly consistent with the slopes 
  presented  in fig.10 of Ref. \cite{pruppacher1995}.

  To estimate the value of  $\gamma$  
  at   the melting point  we extrapolate our data to $\Delta T = 0$  
  (using the averaged slope of -0.25mN/(Km) for the TIP4 family models).
The extrapolations are shown in Fig. \ref{sigma_vs_T} and the values of $\gamma$ at coexistence thus obtained are reported
in Table \ref{melting_properties}. 
   
  Within the TIP4P family the value of $\gamma$ increases with the strength of the hydrogen bond. Therefore 
 within this family one could state that $\gamma$ increases with the melting enthalpy or with the melting point. 
  The correlation between $\gamma$ and the melting enthalpy was first proposed by Turnbull \cite{kelton,turnbull}. 
 Another correlation between $\gamma$ and the melting point has been proposed by Laird \cite{laird_rule}. 
 We indeed confirm that for the TIP4P family both  
 the correlation of Turnbull \cite{turnbull} and Laird \cite{laird_rule} could be useful to predict the trends in $\gamma$. 
 In fact, mW and TIP4P/ICE both have the same melting point and melting 
 enthalpy. According to the Turnbull recipe, or  the Laird recipe, they should have a quite similar value of $\gamma$. 
 This seems to be consistent with the results of this work. 
 
Moreover the results presented in Table \ref{melting_properties} are in reasonable 
 agreement with results obtained by other authors. 
 Using  the cleavage method and averaging over the basal, primary prismatic and secondary prismatic planes, 
   the value of $\gamma$ for TIP4P , TIP4P-Ew\cite{horn04} (a model with similar properties to TIP4P/2005) 
has been reported to be 
 $26.5(4)$, $27.6(5)$ $mJ/m^2$ respectively \cite{gammadavid}.  
Using the mW,  Ref.\cite{galli_mw} estimated $\gamma$ to be $31 mN/m$, in reasonable agreement with our estimate.
 Experimentally, the value of $\gamma$ for the ice Ih-water interface has been reported to be between 
 27 and 35  mN/m. 
  The most cited work is that of Ref.\cite{gamma_exp} which reports a value of $29.1mN/m$. 
  In the absence of  better criteria, we shall assume this to be the most reliable value. According to that, TIP4P/2005 
 provides estimates of $\gamma$ in agreement with experiments, TIP4P being slightly smaller than 
the experimental one, and the value of the mW and TIP4P/ICE slightly higher.

  \subsection{The attachment rate, $f^+$}
  
\begin{figure}[h!]
\includegraphics[width=0.4\textwidth,clip=]{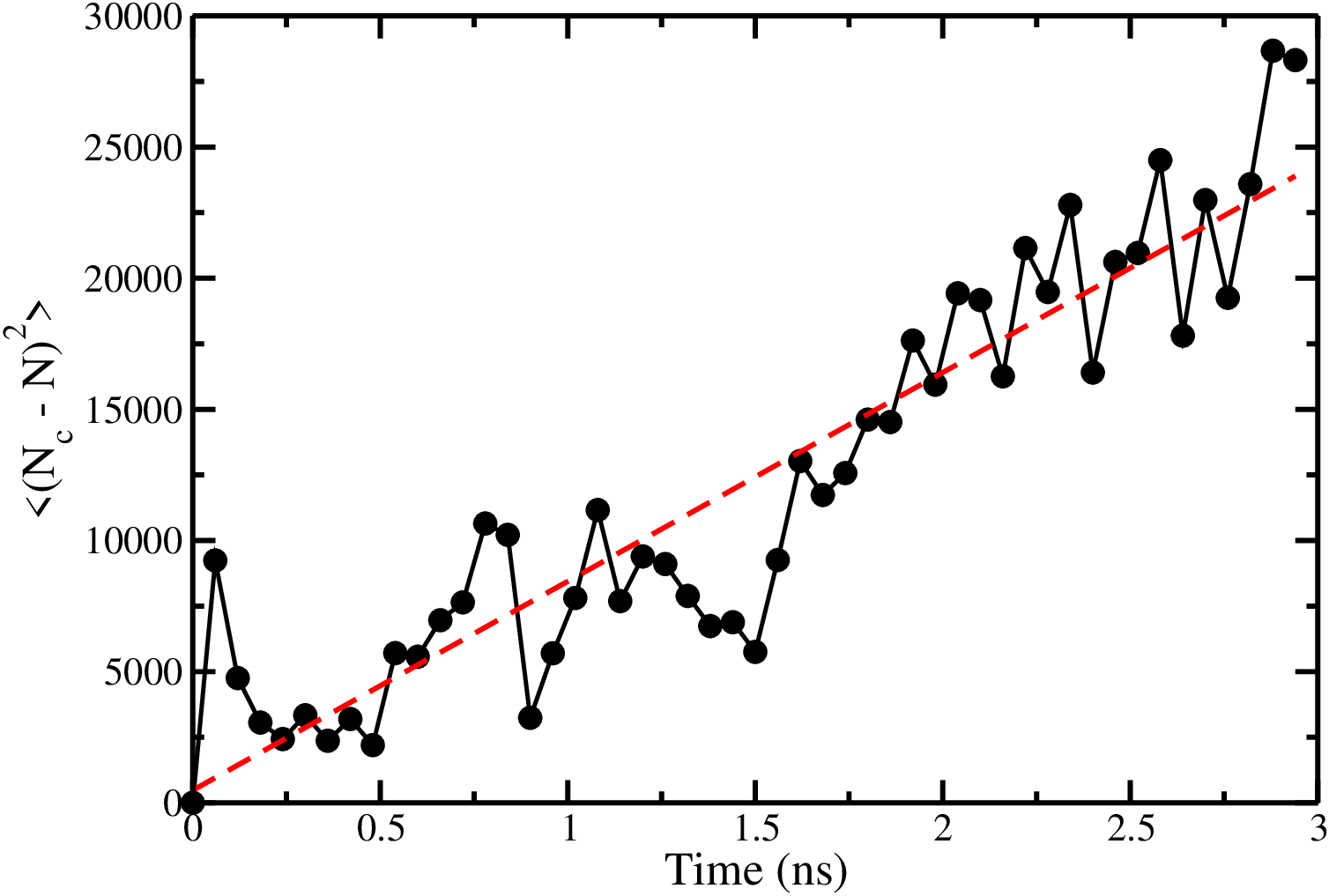}\\
\caption{ Attachment rate for the  cluster of 3170 molecules of the TIP4P model. Results obtained
                from the average of 10 different trajectories. Simulations were performed at 212.5K and 1 bar. Notice that
                in Fig.6 of our previous work\cite{jacs2013}, the results were obtained for the medium cluster of TIP4P/ICE and not for the medium cluster of 
                the TIP4P/2005 model as stated in the caption.} 
\label{attachment_rate}
\end{figure}

  When computing  the attachment rate via Eq. \ref{eqattach}, we observe that the results obtained for TIP4P, TIP4P/2005 and TIP4P/ICE 
 are quite similar. The attachment rate is obtained after  running 10 molecular dynamics trajectories at the temperature that makes the cluster critical (30 trajectories were performed in the case of the mW model ).  
 In Fig. \ref{attachment_rate} we show (for the TIP4P model) the mean squared 
displacement (as obtained from the average of the 10 trajectories) 
 of the cluster size as a function of time for the ice cluster of 3170 molecules (see also the Supplementary material \cite{material_suplementario}).
All trajectories start from the same configuration and differ in the initial set of Maxwellian momenta.
The results of Fig.\ref{attachment_rate} were fitted to a straight line and the attachment rate is just half the value 
of the slope. 
The fact that we are starting all runs from the
same configuration (although with different momenta) may have some impact on the computed slopes 
as pointed out recently by Rozmanov and Kusalik\cite{kusalik_isokinetic}. This can be minimized by excluding the short-time behaviour from the calculation of the attachment rate.
In Table II , we have determined the attachment rate for the medium cluster  using 
both the entire window time and times larger than 1.5ns (results in parenthesis). 
As it can be seen, the impact on the attachment rate is small.

\begin{figure}[h!]
\includegraphics[width=0.4\textwidth,clip=]{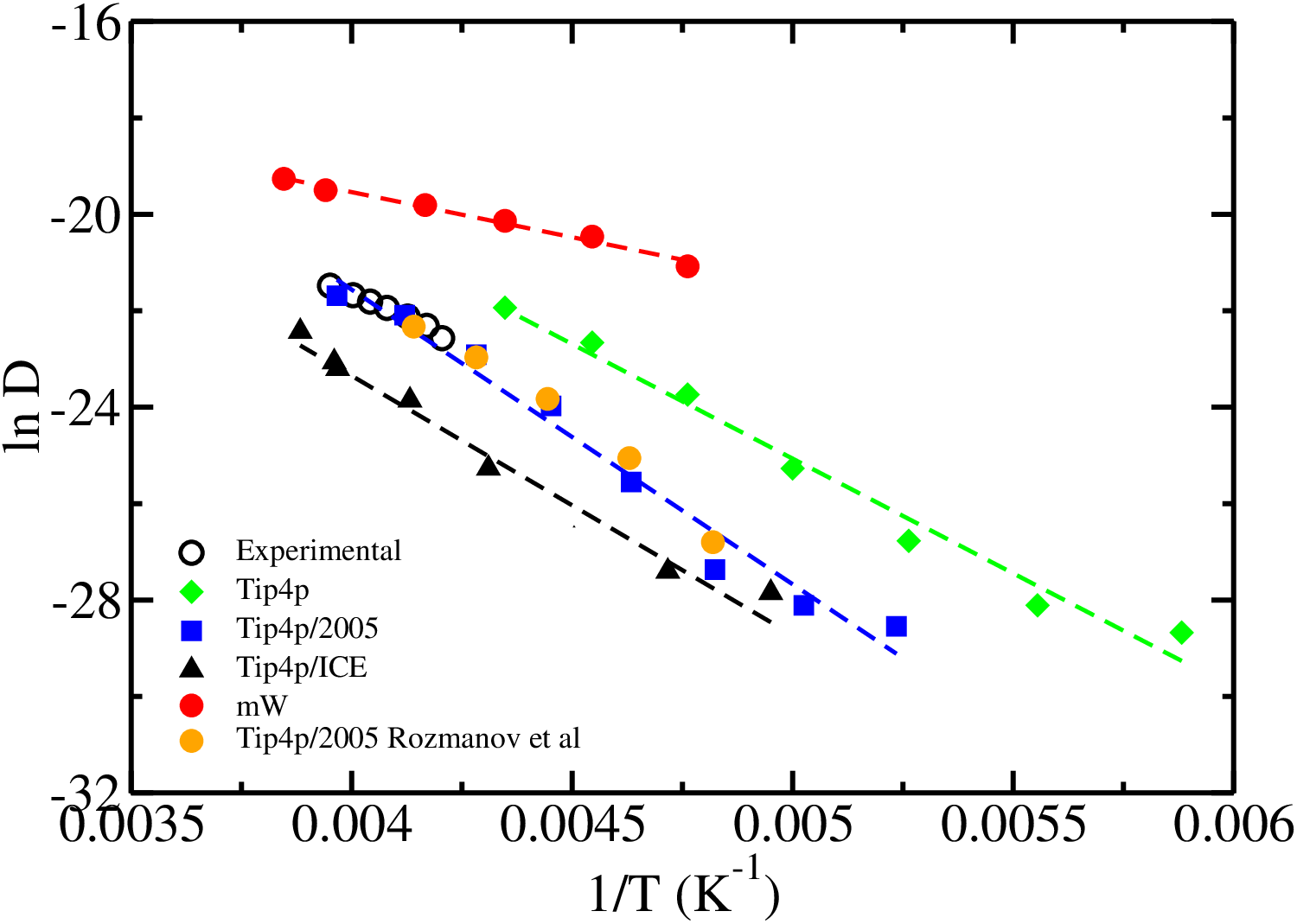}\\
\caption{ The diffusion coefficients for  TIP4P/2005, TIP4P/ICE, TIP4P and mW models. Symbols correspond to  simulation results of this work. 
Lines were obtained from an Arrhenius fit. For the TIP4P/2005 model we have also included the results from Rozmanov and Kusalik\cite{rozmanov12_D} (orange circles) for 
temperatures up to 210K.} Experimental values: open circles\cite{price98}. 
\label{difusion_plot}
\end{figure}

From the slope of the curve shown in Fig. \ref{difusion_plot} one can obtain $f^+$ via Eq. \ref{eqattach}. 
 For the smallest cluster ,  $f^{+}$ is of the order of $10^{11} s^{-1}$ whereas for the largest 
 cluster is of the order of $10^{12} - 10^{13} s^{-1}$.
The results for the attachment rate $f^+$ are shown in Table \ref{jacs_table}. Notice that there was a misprint in 
the main text in our previous work\cite{jacs2013} where we stated that the attachment rate for TIP4P/2005 of the medium 
cluster was 70 $ 10^{9} s^{-1}$. The correct value ( shown in Table \ref{jacs_table}) is 1.2 $\cdot 10^{12} s^{-1}$ and this correct value
was used in the calculations of our previous work \cite{jacs2013} leading to a value of  $log_{10} J$ of $-83$ which is the same as 
that reported here in Table \ref{jacs_table}).  

According to Ref. \cite{kelton}, since the attachment rate $f^{+}$, is related to the time required for a molecule 
to attach to the solid cluster, one could express it as 
\begin{equation}
\label{attachment}
   f^{+} = \frac{ 24 D ( N_{c} )^{2/3} } {\lambda^2 }
\end{equation}  
$N_{c}^{2/3}$   is  the number of molecules at the cluster's   surface and 
$\lambda^2/D$ is the  time required for a molecule to diffuse a given length   $\lambda$ ( D being  
  the diffusion coefficient of the supercooled liquid phase). 
Having numerically computed $D$ at few temperatures, one could use an Arrhenius-like 
expression to estimate the diffusion coefficient as a function of temperature below melting:
\begin{equation}
\label{arr}
  \ln D = \ln D_{0} - \frac{E_{a}}{RT} 
\end{equation}
whose coefficients for each model are  presented in Table \ref{D_fit}. 
\begin{table}[h!]
\caption{Coefficients of the fit of Eq.\ref{arr} to the diffusion coefficient  of supercooled water for the 
TIP4P, TIP4P/2005, TIP4P/ICE  and mW water models. }
\label{D_fit} 
\centerline{
\begin{tabular}{c|ccc}
Model & $\ln(D_0$/($m^2/s$))  & $ E_a/(kJ/mol)$ & \\
\hline
TIP4P & -1.30 & 39.526 \\ 
TIP4P/2005 & 2.88 & 50.803 \\
TIP4P/ICE & -1.84 & 44.709 \\
mW & -13.46 & 12.890 \\
\hline 
\end{tabular}}
\end{table}

Figure \ref{difusion_plot} clearly shows that an  Arrhenius-like expression is 
 sufficient to describe the variation of D with T  for the temperature range considered in this work (i.e from the melting point up 
to temperatures of about 60K below melting).

It is interesting to point out that D does not decrease much with temperature in 
the case of the mW model. The decrease of D with T is more pronounced in the case of the TIP4P potentials.   
In the figure we have also included experimental results \cite{price98}. As  can be seen, the TIP4P/2005 model is able to 
describe the experimental values reasonably well. As shown in Fig.6 our values of D for TIP4P/2005 are entirely 
consistent with those determined previously ( for temperatures up to 210K ) by Rozmanov and Kusalik\cite{rozmanov12_D}.

Having determined the value of D, we can estimate the value of $\lambda$ required to reproduce
 the results of $f^{+}$ obtained in this work using Eq. \ref{attachment} (reported in 
  Table \ref{jacs_table}).  The value of $\lambda$ (see Table \ref{jacs_table}) is of about one molecule diameter (i.e 3.5 $\AA$) 
and does not depend strongly neither on  temperature nor on
 the water model. 
  This means that in order to obtain  fast and reasonable estimates of $f^{+}$ over
a  broad range of temperatures,  
  one  could in principle  only need to determine D, without having to recur to the 
   expensive calculations needed to 
compute   $f^{+}$ using Eq.\ref{attachment}.

  The attachment rate for the mW is about 2-3 orders of magnitude larger than that for the other models.  
  Once again $\lambda$ is of the order of a molecular diameter. 
 The larger value of $f^{+}$ for the mW can be explained by taking into account the fact that for this model D is much larger
 than for the rest of the models ( and for real water) corresponding to an enhanced dynamics. 
 Therefore we should point out that both $f^{+}$ and D decreases with T much less in the mW model than in other models. 
 The absence of explicit hydrogens provokes higher values of D, $f^{+}$ and faster dynamics. 
  If the nucleation free-energy barrier   of this model is similar to that of the other models considered in this work, then by considering the kinetic 
 prefactor one should expect the kinetics of this model being three orders of magnitude faster than that of the 
 other models.

 \subsection{The kinetic prefactor }
 
 The kinetic prefactor required to estimate J is given by the product of $\rho_f$, Z and $f^+$. 
The number density of the liquid, of the order of $10^{28}$ molecules/m$^3$, 
does not change much with temperature. 
The product Z$f^+$ does not have a strong temperature dependence either given that 
as the temperature decreases Z increases and $f^+$ decreases.
Thus, we find that Z$f^+$ is of the order of $10^9$ $s^{-1}$ for the TIP4P family of models.
Hence, the kinetic prefactor for TIP4P-like models is 
of the order of $10^{37}$ m$^{-3}$s$^{-1}$.

As TIP4P/2005 describes quite well the diffusion 
coefficient of water at different temperatures, we believe that this is the order of magnitude of the 
kinetic prefactor of real water. 
Notice that for the mW model the kinetic prefactor is 2-3 orders of magnitude larger that 
for the TIP4P models. Therefore for the mW model the kinetic prefactor is of the order of $10^{40}$ m$^{-3}$s$^{-1}$. 

 \subsection{The free-energy barrier, $\Delta G_c$}
 
 The free-energy barriers for all clusters considered in this work are reported in Table \ref{jacs_table}. 
 
For the largest clusters the free energy barrier is about 500$k_B$T , for the medium clusters about 
250$k_B$T and for the smallest clusters  about 80$k_B$T and the 
differences among models are not  particularly large. 
The lowest value of the free-energy barrier corresponds to TIP4P and the largest to 
mW although the differences are not too large. 
For the mW model, which has a somewhat larger value of $\gamma$, one would expect the largest 
free-energy barriers. However, this is not the case given that  both the ice density and  $\Delta \mu$ 
are very large, partially  compensating  this effect. 

For  TIP4P/ICE our results differ from those of Ref.\cite{geiger_dellago}, 
where by means of umbrella sampling, a free-energy barrier of 
 35$k_{B}T$ and a critical cluster  of 300 molecules at a temperature of 235K was reported. Our estimate is of 80$k_{B}T$ and
 600 molecules at the same temperature. Performing  10 independent runs  starting from an initial configuration of a 300 molecule cluster, we
 observed that the cluster always melted  after 30-50ns. These results suggest that
a cluster of 300 ice molecules is most likely sub-critical for these thermodynamic conditions.
Although the order parameter used in Ref.\cite{geiger_dellago} is different from that used in this work we found that both 
criteria differ only in about ten per cent in identifying the size of a given cluster. 

We have included in Table \ref{jacs_table} the statistical error in  $\Delta G_c$.  Once the order parameter is chosen 
then we can determine $N_c$ accurately ( so that there is practically no statistical error in the determination of $N_c$).     
We have an uncertainty of about 2.5K in $\Delta T$ , and that provokes an uncertainty of about 7\% in both $\Delta \mu$ and $\gamma$. 
Notice that these two errors are not independent since we are obtaining $\gamma$ from Eq. \ref{ncrit}. Therefore if $\Delta \mu$ is underestimated by
7\%  then $\gamma$ will be underestimated by  7\% also. According to this the statistical error in  $\Delta G_c$ is also about 7\%. 
The statistical error for $\Delta G_c$ is shown in Table \ref{jacs_table}. This statistical error can be reduced by performing 
 more trajectories. In principle, this statistical error can be reduced at the expense of using a huge 
amount of CPU time.

There is however an additional source of uncertainty which is systematic and can not be reduced by performing more trajectories. 
Different order parameters will yield  somewhat different values of $N_c$ (mainly due to the interfacial region). It is difficult 
to evaluate the impact of this systematic error (in fact if you know exactly the magnitude and sign of the systematic error you can 
always correct your results to the exact value !) and for this reason we shall just provide a rough estimate. 
Different (reasonable) order parameters 
gave differences 
of up to $N_{c}^{2/3}$ molecules for $N_c$. This gives a systematic error in $N_c$ of about  5\%, 7\% and 12\% for the large, the medium and the small cluster respectively. 
of about  10\%. It follows then, that this systematic source of error would affect the values of $\gamma$ \ref{ncrit} by about 5/3\%,7/3\% and 4\% 
respectively. These systematic errors are smaller than the statistical error for $\gamma$ (of about 7\%). 
Since $\Delta G_c$ scales with $\gamma^3$ (see Eq. \ref{eq_G_cnt})then the systematic error would affect
the values of $\Delta G_c$ by about 5\% 7\% and 12\% respectively. We mentioned previously that the stochastic error in $\Delta G_c$ is of
about 7\%. It seems that the systematic error for $\Delta G_c$ is similar to the stochastic error.
In Table \ref{jacs_table} we have included only the statistical errors in $\Delta G_c$ . If one wishes to estimate the total error (i.e 
including the systematic error) one can roughly multiply the error of Table  \ref{jacs_table} by two.

\subsection {The nucleation rate, J}

The homogeneous nucleation rate J is defined as the 
number of critical nuclei per unit of volume and time. 
Results of the nucleation rate are also reported in Table \ref{jacs_table}, where we conclude that 
the order of magnitude changes from $10^{-180}$m$^{-3}$s$^{-1}$ for the temperatures around 15 degrees
below melting to about $10^{0}$m$^{-3}$s$^{-1}$ for temperatures about 35K below.

Due to the number of approximations we used to 
determine these numbers 
one might wonder whether our predictions for the nucleation rate J are reliable or not.
The statistical error in $log_{10} J$ is presented in  Table \ref{jacs_table}. The error in the kinetic prefactor
in the expression of J has an error of about one order of magnitude. From the  error in $\Delta G_c$ it is easy to 
obtain its contribution to the error in $log_{10} J$ simply by dividing by 2.3 ( from the conversion natural to decimal 
logarithms). Therefore the total statistical error in  $log_{10} J$  is obtained after adding these two terms. 
As discussed previously , if systematic errors were also included then the error in  $log_{10} J$ presented in Table \ref{jacs_table} should
be (roughly) multiplied by two. From this it follows that the total error in $log_{10} J$ (stochastic and systematic) is of about 
40, 20 and 6 for the largest, medium and smallest clusters considered in this work.

We  use the results obtained at three different temperatures to estimate J over a broad range of temperatures. For this purpose 
we need to calculate the height of the nucleation free energy barrier, $\Delta G_c$, and 
the kinetic prefactor, $Z \rho_f f^+$, for any temperature and obtain the rate with Eq. 
\ref{eqrate}. 
To obtain $\Delta G_c(T)$ we use Eq. \ref{eq_G_cnt}, where the functions $\gamma(T)$, $\Delta \mu(T)$ 
and $\rho_s(T)$ are required. 
For $\gamma$, we assume that it changes linearly with T in the way shown 
by the fits in Fig. \ref{sigma_vs_T} (dashed lines). The chemical potential difference as a function of
temperature is calculated by thermodynamic integration (see Fig. \ref{delta_mu}). 
The density of the solid as a function of temperature is taken from a linear fit to the results
of Table \ref{jacs_table}.  
To obtain the kinetic prefactor as a function of temperature we need $\rho_f(T)$, Z(T) and $f^+(T)$.  
The density of the fluid changes smoothly with temperature and we have considered a constant value 
of 0.94 g/cm$^3$ for all models. By using Eqs. \ref{zeldovich} and \ref{ncrit} and with the functions $\gamma(T)$
and $\Delta \mu(T)$ above described one can easily obtain Z(T). Finally, we use Eq. \ref{attachment} to obtain $f^+(T)$.
Eq. \ref{attachment} requires, in turn, D(T) and $\lambda(T)$. For D(T) we use the fit given by Eq. \ref{arr}. 
For $\lambda$ we take a value independent of temperature and equal to an average between the values found for the three clusters 
(for all cases $\lambda$ is of the order of a molecular diameter).    
With these approximations (which appear quite reasonable
after the results presented so far) we can obtain J for any
value of $\Delta T$ (supercooling). 

\begin{figure}[h!]
\includegraphics[width=0.4\textwidth,clip=]{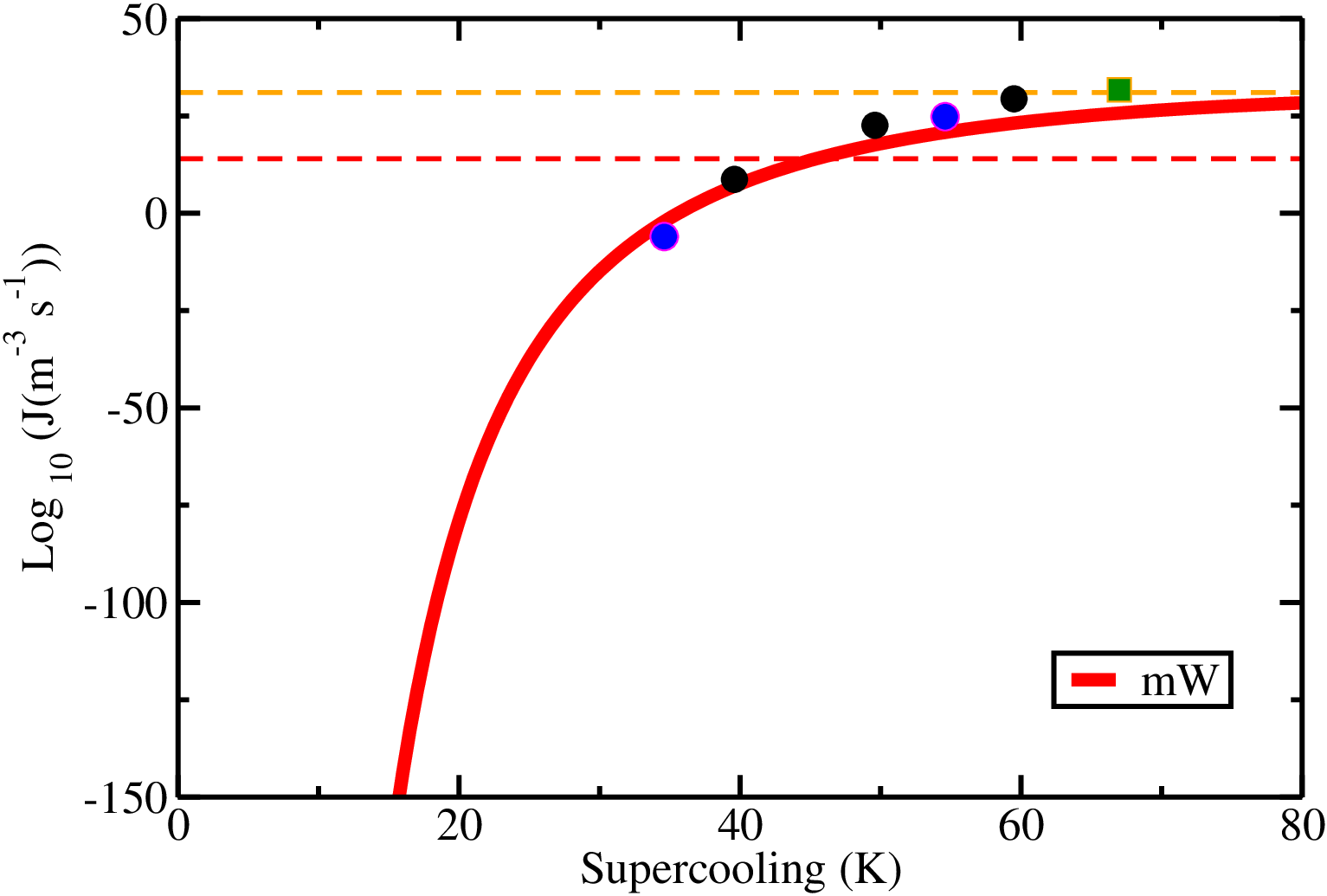} a)\\
\includegraphics[width=0.4\textwidth,clip=]{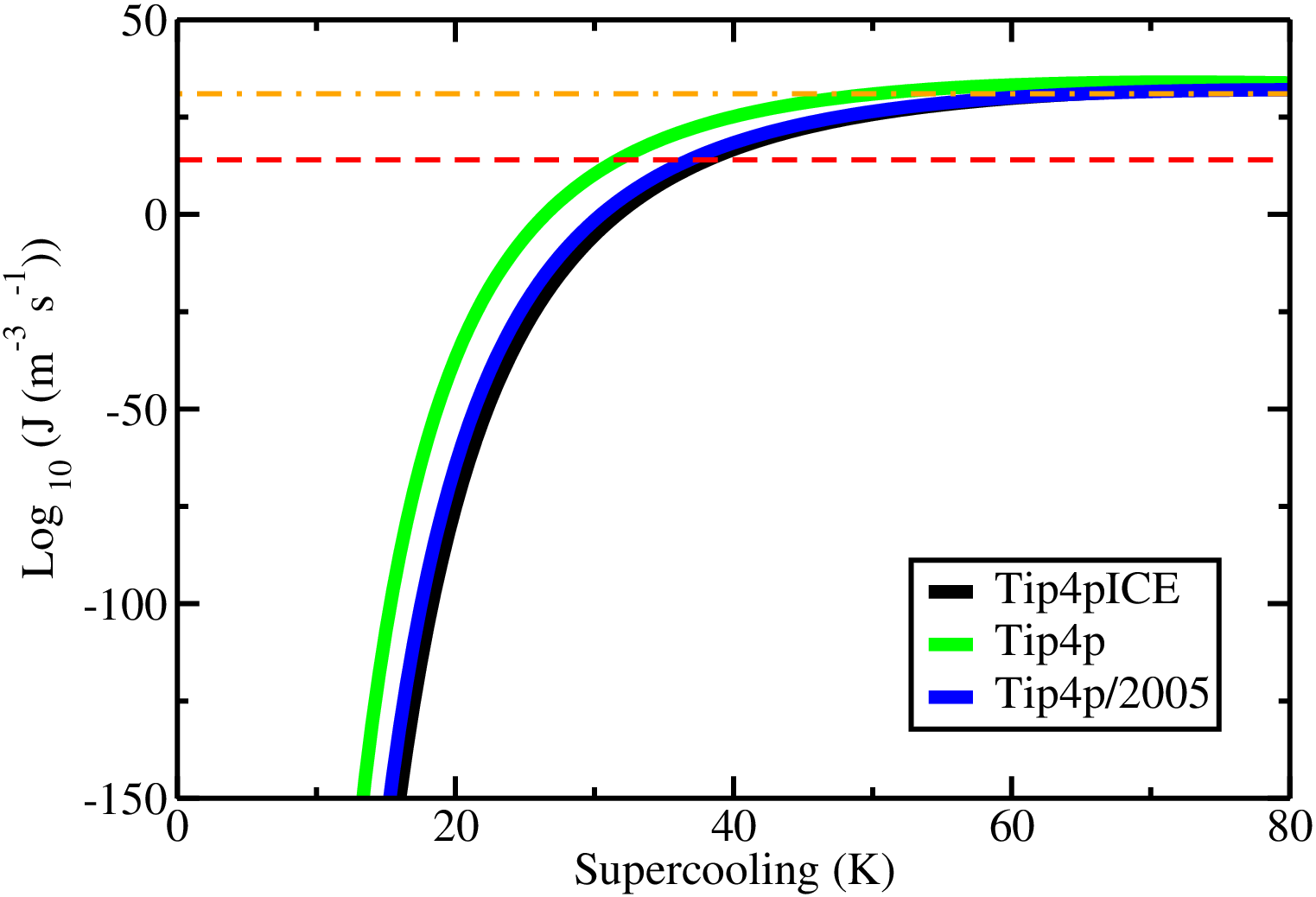} b)\\
\includegraphics[width=0.4\textwidth,clip=]{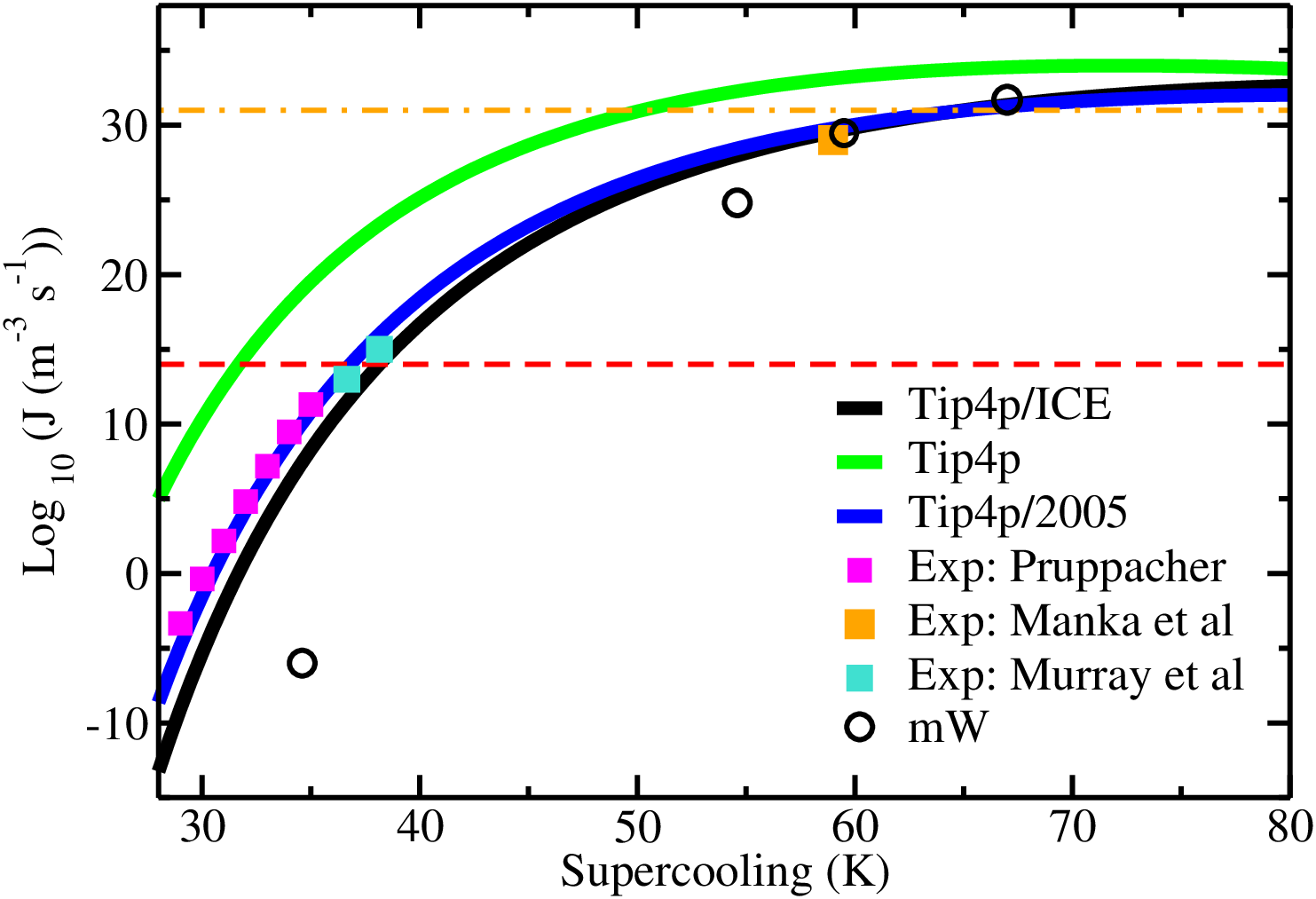} c)
\caption{
Values of J for several water models, as obtained in this work, from experiment, and in previous work ( in the case of the mW model). 
The horizontal lines correspond to $log_{10} J/(m^{-3} s^{-1}) =14$ and
 $log_{10} J/(m^{-3} s^{-1}) =31$ which are the approximate values of J at the homogeneous nucleation temperature in experiments and in simulations respectively. 
a) J for the mW  model as obtained in this work ( red solid line) . Blue circles are results
at 240K and 220K from Ref. \cite{galli_mw}, black circles are the results from Russo et al. \cite{russo14} 
and green square at 208K from Ref. \cite{nature_valeria}. 
b) J for the TIP4P , TIP4P/ICE and TIP4P/2005 models.
c) J of the models studied in this work (solid lines) compared to experiments (filled squares) of  Pruppacher \cite{pruppacher1995}, Murray et al. \cite{murray2010kk} and Manka et al \cite{manka2012}. Empty circles are estimates of J for the mW model as reported in References \cite{galli_mw,nature_valeria,russo14}.
Notice that, in the c panel, both x and y-axis  differ from the other two panels. }
\label{J_plot}
\end{figure}

In fig.\ref{J_plot}, we present the  results of the logarithm of J as a function of the degree of supercooling
for different water models.
In figure \ref{J_plot} (a) we
show our results for the mW model and  compare it with previous calculations of J. 
At 240K our value of J is about 4 orders of magnitude higher that the value reported 
by Li et al.\cite{galli_mw}, whereas at 220K , 215K and 208K our value is about 4-6 orders of magnitude lower than the values reported by 
Li et al.\cite{galli_mw}  (220K using forward flux sampling), Russo et al.\cite{russo14} (215K using umbrella sampling) 
and Moore and Molinero\cite{nature_valeria} (at 208K using brute force simulations).
From this, we conclude that our predictions of the nucleation rate for the mW model are 
in reasonably good agreement with results previously reported in the literature, 
taking into account that the approach used here is an approximate one and that the uncertainty in J 
from our technique is about 6 orders of magnitude at high supercooling (coming from the uncertainty in determining
the temperature at which the cluster is critical and the procedure used to distinguish solid from liquid-like molecules). 
We estimated the size of the critical cluster to be of 86 molecules at 205K for the mW model, in excellent agreement with the value reported 
by Moore and Molinero for the same model and temperature which was of about 90 molecules\cite{nature_valeria}. 
Since Russo et al.\cite{russo14} have
determined not only nucleation rates but also, $f^{+}$ , $N_c$ and the free energy barrier it is interesting to have a closer comparison term by term. 
This is done in Table \ref{mw_a_fondo}. As  can be seen, the agreement for all individual terms is quite good.
For $log_{10}(J)$ is reasonable taking into account that, as discussed previously, our uncertainty in 
$log_{10}(J)$ at high supercoolings, when all possible sources of error are considered, is of about 6 orders of magnitude.

\begin{table}
\caption{Contributions  (term by term) to J for the mW model as obtained in this work and as obtained by Russo et al.\cite{russo14}. Results of 
this work were obtained at $p=1bar$ whereas those of Russo et al.\cite{russo14}  were obtained at $p=0bar$. This small difference of pressure is not expected to
affect any of the terms of the table. $f^{+}$ is given in $s^{-1}$, $\Delta G_c$ in $k_BT$ units and J 
in $(m^{-3}s^{-1}$) } 
\centerline{
\begin{tabular}{ccccccc}
\hline
Source  & T(K) &  $N_c$ &  $f^+$  & Z  &  $\Delta G_c$  &  $log_{10}(J)$ \\
\hline
\hline
This work   & 215.1 & 128 & 1.0 $\times 10^{13}$ & 0.016 & 38.2 & 23.2 \\
Russo et al & 215.1 & 81 & 0.8 $\times 10^{13}$ & 0.018 & 23.5 & 29.4 \\
\hline
This work   & 225   & 213 & 2.0 $\times 10^{13}$ & 0.0109 & 51.5 & 17.5 \\
Russo et al & 225   & 180 & 2.6 $\times 10^{13}$ & 0.0115 & 40.1 & 22.58 \\
\hline
This work   & 235   & 405 & 4.1 $\times 10^{13}$ & 0.0070 & 76.3 & 6.8 \\
Russo et al & 235   & 400 & 4.7 $\times 10^{13}$ & 0.0077 & 72.0 & 8.7 \\
\hline
\label{mw_a_fondo} 
\end{tabular}}
\end{table}

Let us now describe the results for TIP4P like models. 
As it can be observed in figure \ref{J_plot} b), the values of homogeneous nucleation rates for  
TIP4P/2005 almost coincide (to within the error bars) with those computed for 
TIP4P/ICE, whereas the ones for TIP4P are  slightly higher. 
In Fig. \ref{J_plot} c) the values of homogeneous nucleation rates for  TIP4P/2005 are 
compared to experimental ones at the temperatures where most experiments are available 
(i.e between 235K and 240K). From the data we conclude that the results of TIP4P/2005 are consistent 
with the experimental ones ( taking into account the combined uncertainty of both experimental and
simulation results). 
Thus it seems that TIP4P/2005 is able to reproduce not only  
the ice density, the ice-water interfacial free-energy and the diffusion coefficient, but also the nucleation rate J of real water. 
We also compare our results with  recent experimental work where homogeneous ice nucleation was 
measured in nanoscopic water droplets \cite{manka2012}. By using such small droplets
in Ref. \cite{manka2012} homogeneous ice nucleation was probed at an extremely high supercooling
(59 K below melting). Notably, the agreement between the TIP4P/2005 model and the experiments
of Ref. \cite{manka2012} is also very good. As  can be seen in fig. \ref{J_plot} c) , for the mW model, the 
values of J obtained in previous works\cite{galli_mw,nature_valeria,russo14} seem
to be in good agreement with the experimental results when the supercooling is large. In fact the agreement 
with the recent results of Manka et al. \cite{manka2012} is quite good. However for moderate 
supercooling J of the mW model seems to be lower than those found in experiments, this is most likely 
due to the high value of the interfacial free energy $\gamma$ of the model.

  Another interesting feature is that for the TIP4P model the nucleation rate reaches a maximum value and after that 
  it  decreases slightly (see Fig. \ref{J_plot} c) . For the other TIP4P models one may expect  similar behavior but 
  at lower temperatures.  The maximum is caused by the fact that  
     the thermodynamic driving force for nucleation increases as the temperature decreases (i.e the free-energy barrier decreases) 
     and at the same time   the kinetic prefactor decreases dramatically with temperature and at very low temperatures 
  becomes the dominant factor.  The fact that J may reach a maximum has been already suggested 
  by Jeffery and Austin \cite{jeffery1997} and is consistent with the experimental results of Refs. \cite{huang_bartell,bartell94} 
  when studying the freezing of water clusters at very low temperatures (i.e 72K below melting), although
  is not entirely clear if at this high supercooling the formation of ice is limited by ice nucleation or 
  by growth (see discussion below) .

\subsection{Homogeneous nucleation temperature}

The homogeneous nucleation temperature $T_H$   is a kinetic concept. 
 $T_H$  is the temperature below which water does not exist in its liquid phase 
 (because it  freezes).  However, to properly define $T_H$, we need to 
specify both the sample size and the duration of the experiment. 
  The experimental value of homogeneous nucleation temperature ($T_H^{exp}=235K$)
can be approximated by the  temperature at which one critical ice cluster is formed in a 
spherical micrometer-size  water droplet and for one minute: 
\begin{equation}
     J =  \frac{1}{ \frac{4}{3} \pi  (2\;\; 10^{-6})^3    60 }   =   10^{14}/(m^{3} s).  
\end{equation}
This experimental rate  
is represented by a dashed line in figure \ref{J_plot}.
As  can be seen in figure \ref{J_plot} b) in the case of TIP4P/ICE and TIP4P/2005, $T_H^{exp}$ is 
 located about 37K below melting, in reasonable agreement with the experimental value. 
In the case of TIP4P $T_H^{exp}$ is slightly lower 
(around 30K below melting).

Let us now estimate the free-energy barrier height when $J=10^{14}/(m^3 s)$. For the TIP4P models it is of the order of  53$k_B$T 
(given that the kinetic prefactor is about $10^{37}$ $m^{-3}s^{-1}$) whereas for the mW model it is of 
the order of  60$k_B$T. Since the values of J for TIP4P/2005 agree quite well with experiments, this strongly suggest that 
at the experimental value of $T_H^{exp}$ (i.e 235K) the free energy barrier for nucleation is  about 53$k_B$T. 
 It is interesting to point out that the 
 attachment rate $f^{+}$, of the mW model, is of the same order of magnitude of that found for 
 LJ  systems. 
 Therefore for systems formed by atoms/ions , $f^{+}$ seems to be of the same order magnitude.
  Obviously for these systems the free energy barrier must be  about 60$k_B$T at $T_{H}^{exp}$ \cite{kelton}. However, for 
 water, $f^{+}$ is three orders of magnitude smaller and the free energy barrier at $T_{H}^{exp}$ must be  
 about 53$k_B$T.  In other words  as a rule of thumb one can state that the experimental homogeneous 
 nucleation of water in micrometric droplets is the temperature at which the free energy barrier becomes of about 53$k_BT$. 
It is interesting to point out that both the value of the homogeneous nucleation temperature  and of associated free energy barrier depend
on the volume of the droplets with which the experiments are performed. The considerations above are all for 
micrometric  water droplets, which is the most widespread experimental set up for the study of homogeneous ice nucleation.
But this is not always the case.  
In fact, in a recent work, by using nanoscopic droplets much higher nucleation rates, and smaller nucleation
barriers, were probed \cite{manka2012}. Therefore, the so called ``homogeneous nucleation line'' depends on the volume of
the water droplets and should not be taken as a definite limit for the existence of supercooled liquid water.  

When dealing with computer simulations, both length and time-scales are quite 
different. The simulation value of the homogeneous nucleation temperature 
($T_H^{sim}$) can be estimated as the temperature  at which one critical ice cluster is formed in a 
simulation with a box side of $40 \AA$ (corresponding to a typical supercooled water density 
of about $0.94 g/cm^3$ in a system of 2000 molecules) for 1 microsecond.
At these conditions, the nucleation rate takes the  value:
\begin{equation}
    J =   \frac{1} {  (40\;10^{-10})^{3}   10^{-6} }  =  10^{31}  /(m^{3} s) 
\end{equation}
as represented by a dashed line in figure \ref{J_plot}.
As  can be seen in figure \ref{J_plot} b) in the case of TIP4P/ICE and TIP4P/2005, $T_H^{sim}$ is 
 located about 60-65K below melting. Whereas  once more in the case of TIP4P $T_H^{exp}$ is slightly lower 
(around 50 below melting).
Again, knowing the nucleation rate, one could estimate the free-energy barrier 
height at $T_H^{sim}$  for  TIP4P/ICE, TIP4P/2005 and TIP4P to be of the order of  13$k_B$T. 
Whereas the free-energy barrier  height at $T_H^{sim}$ for mW is of the order of  19$k_B$T
(given that the kinetic prefactor for this water model is three orders of magnitude larger).
When simulating simple/atomic fluids (such as hard spheres \cite{hs_filion}, Lennard-Jones\cite{baidakov:234505} or NaCl\cite{nacl_valeriani}), 
spontaneous nucleation within reasonable time-scale can be observed with brute force simulations when the free-energy barrier height is 
of the order of 18$k_B$T.
  
  \subsection{Growth rate and Avrami's law}

\begin{figure}[h!]
\includegraphics[width=0.4\textwidth,clip=]{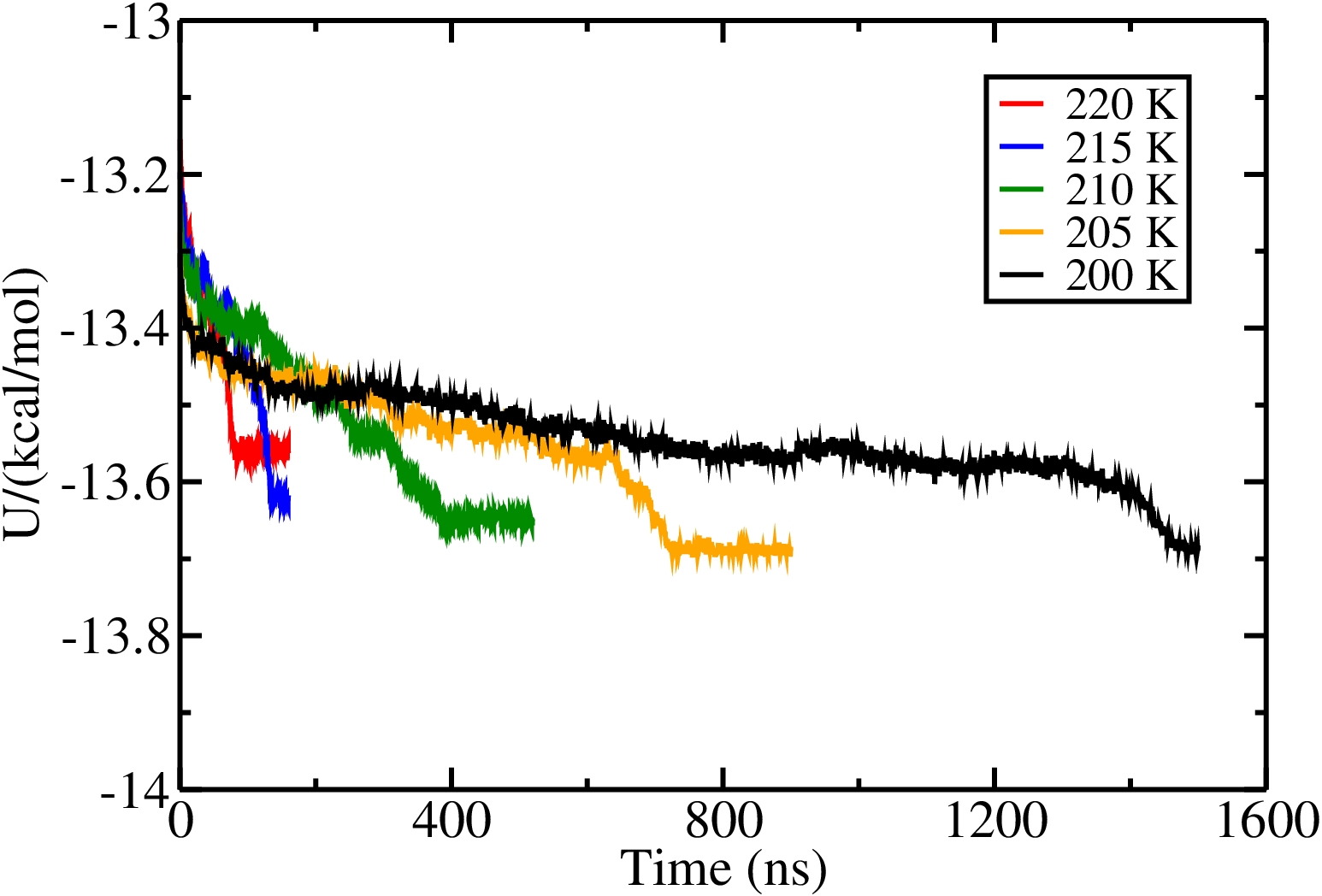}\\
\caption{ Evolution of the potential energy with time in direct coexistence runs of the TIP4P/2005 model 
using a slab with 2048 molecules (half ice Ih and half liquid water) at the temperatures, from the left to the right, of 
220K, 215K, 210K, 205K and 200K. }
\label{growth_rate_plot}
\end{figure}

 Rozmanov and Kusalik\cite{kusalik_2011_growth}  have determined the growth rate of TIP4P/2005 for temperatures between the melting point 
 and 210K and fitted their results to a Wilson-Frenkel like expression\cite{wilson00,frenkel32,broughton82}. To estimate the ice growth rate 
  at lower temperatures we have performed direct coexistence simulations for the 
  TIP4P/2005 at 1bar and temperatures below 210K. The system consists of 2048 molecules. In the initial configuration 
  half of the molecules are forming ice and the other half supercooled water (i.e approximately we have 
  a 35 \AA\ layer of ice and a 35 \AA\ layer of water). 
  The evolution of the potential energy with time is shown in fig.\ref{growth_rate_plot}. For all considered temperatures 
  the system freezes completely ( as shown from the final plateau of the energy, from visual inspection of the final configuration 
  and from the analysis of the sample using order parameters). 
  Obviously the time required to form ice is much longer at 200K (1500ns) than at 220K (80ns). 
  To estimate the growth rate we simply divided 35 \AA\ (i.e the thickness of the liquid slab) 
  by the time required to freeze the system.
  Notice that this is used just to provide a rough estimate of the growth rate of ice. A rigorous determination of the 
  growth rate requires performing the analysis over a larger number of independent trajectories. 
  For the three highest temperatures 220K, 215K and 210K the growth 
  rate estimated in this work is fully consistent with that obtained previously by Rozmanov and Kusalik.\cite{kusalik_2011_growth,Kusalik_anisotropy}
  For the two lowest temperatures, the growth rate estimated in this work , 0.049 \AA\ /ns at 205K and 0.025 \AA\ /ns at
  200K , should  be compared to the values  0.056 \AA/ns and 0.040 \AA/ns obtained from the fit of Rozmanov and Kusalik (for the average
  of the different planes).\cite{Kusalik_anisotropy}  Since the agreement is satisfactory we shall assume that the fit of Rozmanov and Kusalik for the ice growth rate, can 
  be used for temperatures below 210K.

In general, nucleation is the limiting step for supercooled liquid water to transform into  ice. 
Therefore, once a critical cluster is formed, ice crystal growth tends to occur quite rapidly. 
However, at low temperatures this might not be the case, since the ice growth rate , u , might be very small. 
 When the growth rate is  small, 
 one should introduce a new parameter,  $\tau_{x}$ (which is the time 
 required to crystallize a certain volume fraction of the sample $\phi$). 
 $\tau_{x}$  depends on two  properties, the 
 value of J,and the value of the growth rate of ice ,  u. 
 The Avrami's equation has been 
 considered for obtaining $\tau_{x}$ \cite{avrami_law,nature_valeria,debenedettibook}. In Debenedetti's book the expression for $\tau_{x}$ is 
 provided and it is given by :

\begin{equation}
    \tau_{x}^{Avrami} =  ( (3 \phi)      / ( \pi J u^3) ) ^{1/4}   
\end{equation}

\begin{figure}
\includegraphics[width=0.4\textwidth,clip=]{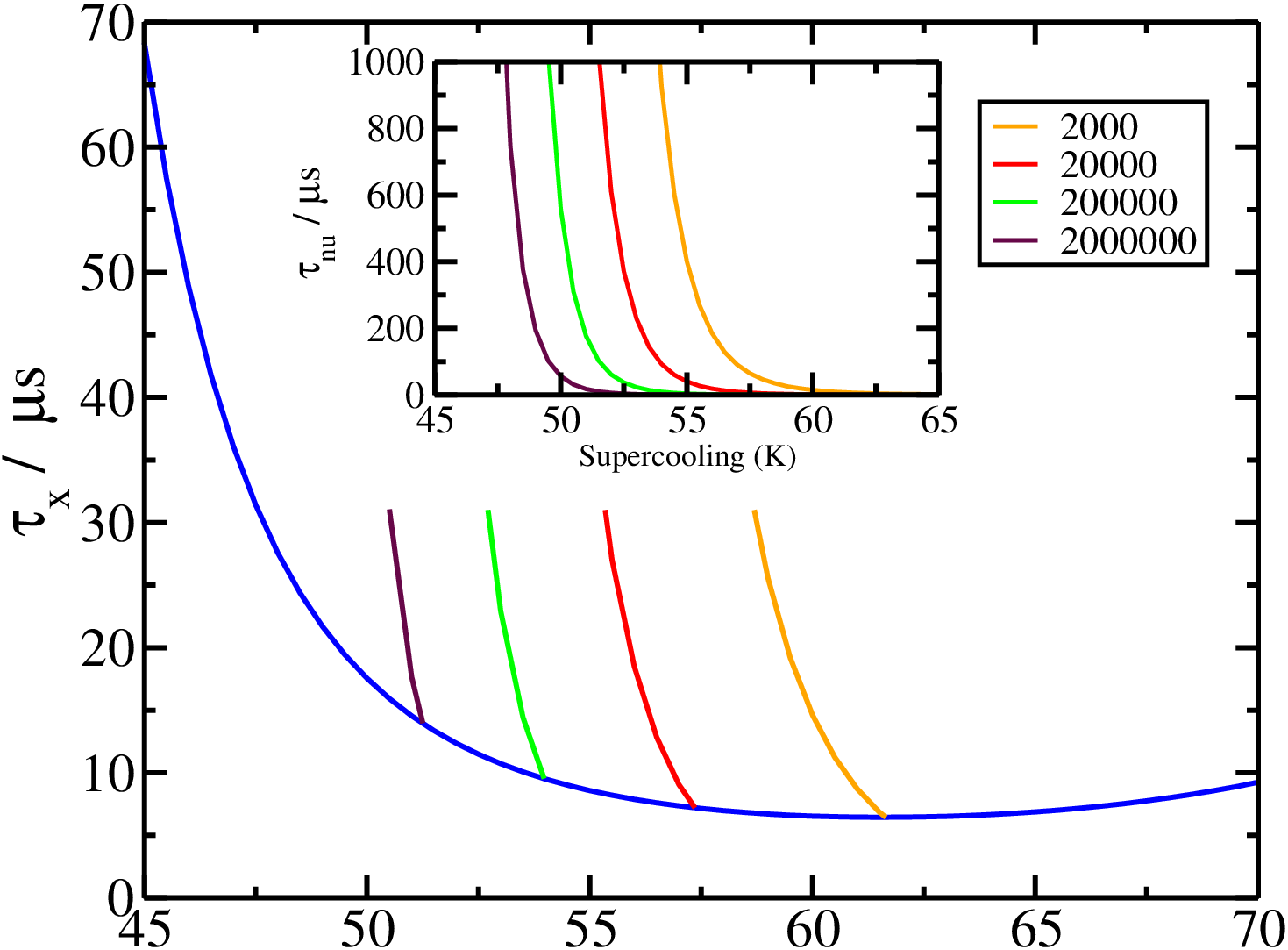} a) \\
\includegraphics[width=0.4\textwidth,clip=]{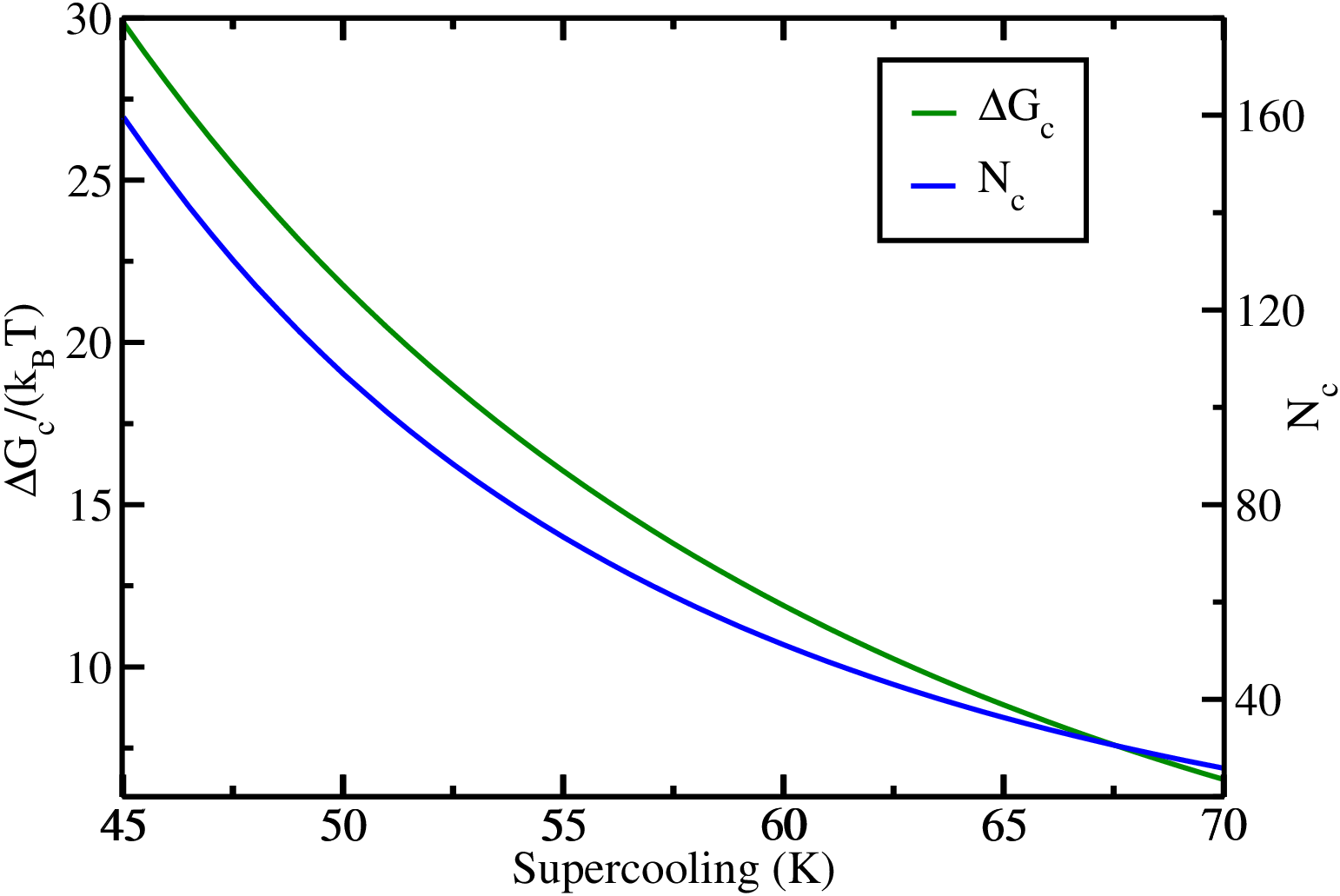} b) \\
\caption{(a): $\tau_{x}$ for $\phi=0.7$ for the TIP4P/2005 model as a function of the supercooling. $\tau_{x}$ is the time necessary 
to crystallize 70$\%$ of the system. Inset: plot of the nucleation time, $\tau_{nu}$, versus the supercooling. (b): 
Free energy barrier for nucleation and size of the critical cluster for TIP4P/2005 as a function of the supercooling.}
\label{tau_plot}
\end{figure}

 By using Avrami's  expression we have plotted $\tau_{x}^{Avrami}$ in fig. \ref{tau_plot}  as a function of the degree of supercooling 
 for the TIP4P/2005 model. Although we used $\phi=0.7$ , $\tau_{x}^{Avrami}$ is practically the same for any value 
 between 0.6 and 0.9 chosen for $\phi$, as $\tau_{x}$ changes as $\phi^{1/4}$. 
As  can be seen the minimum $\tau_{x}$ is of the order of several 
 microseconds. The minimum in $\tau_{x}$ occurs at smaller values of supercooling that the maximum 
 in J. In any case it is important to recall the fact that $\tau_{x}$ rather than J is the 
 relevant magnitude at large supercoolings as the growth rate of ice can be the limiting factor. 
There is still a subtle issue with respect to the application of Avrami's expression.  
 Notice that Avrami's expression contains only the intensive parameters $\phi$, J, and u, so  $\tau_x$ does not
 depend on the size of the system. However, as pointed out by Berg \cite{berg_avrami_finite}, there are important system size effects on $\tau_{x}$ 
 specially when one goes down to the system size typical of computer simulations. 
When the nucleation time $\tau_{nu}$ (i.e  the time required to form a critical nucleus) :

\begin{equation}
     \tau_{nu} =    1/ ( J V )         
\end{equation}

is larger than the diffusive time one can not find a critical cluster
growing in the system until the nucleation time has elapsed. In such regime, Avrami's traditional expression 
can not be applied and the crystallization time is dominated by the nucleation time, that is inversely
proportional to the system's volume.  
Following Berg \cite{berg_avrami_finite}, let us define a parameter $q$ as the ratio of two times, 
 the growth time $\tau_{growth}$ ( i.e the time required to crystallize completely the simulation box after a post-critical nucleus has been
formed) and $\tau_{nu}$ as :
 
\begin{equation}
     \tau_{growth} =   L / u        
\end{equation}

\begin{equation}
     q =   \tau_{growth} / \tau_{nu} 
\end{equation}

 where L is the dimension of one of the sizes of the cubic simulation box.
Notice that q depends on the system size so it is not an intensive property.
In fact for any temperature q tends to $\infty$ as one increases the system 
size to the thermodynamic limit ( the numerator scales as $L$ whereas the denominator scales as $L^{-3}$). 
 According to Berg \cite{berg_avrami_finite} $\tau_x$ can be expressed as :

\begin{equation}
     \tau_x =  \tau_{nu} ( 1 + f_{d}(q) )  
\end{equation}

The function $f_{d}(q)$ behaves as $A_d q$ for values of $q$ smaller than one (i.e when the growth time is smaller 
than the nucleation time) and behaves as $B_d q^{3/4}$ for values of $q$ larger than $64$ (i.e when the growth time is 
larger than the nucleation time). Between these two values one has a crossover behavior. According to this
for values of $q$ larger than $64$ , one recovers the traditional Avrami's expression.
However for small values of q $\tau_{x}$ can be approximated quite well by $\tau_{nu}$.  
Since $\tau_{nu}$ depends on the system size, so does $\tau_{x}$. 
In fig. \ref{tau_plot} we have also plotted the value of $\tau_{nu}$ for system sizes of 2000, 20000, 200000 
and 2000000 molecules of water. For each system size $\tau_{x}$ is given by $\tau_{nu}$ up to the temperature
at which $\tau_{nu}$ intersects Avrami's expression. Obviously as one moves to larger system sizes the intersect moves to 
lower supercooling (i.e higher temperatures) and in the thermodynamic limit, Avrami's expression is valid for all temperatures. However
this is certainly not the case for finite size systems. Notice also that due to the finite size effects
small systems gain an extra stability with respect to freezing (i.e more time is needed to 
freeze the system). One could state that crystallization is controlled by nucleation when $\tau_{nu}$ is much larger 
than $\tau_{growth}$ , and by ice growth when the opposite is true. This behavior is sensitive to the system
size used in the simulations.
 
After the previous discussion it is clear that the conditions where spontaneous crystallization 
of TIP4P/2005 water
could be most affordable in terms of CPU time would be a system of about 20000 molecules at about 195K (i.e 57K below the melting point).
Under these conditions Avrami's expression is valid and from our estimates it should take about 6 microseconds
to freeze the system. 
That may explain why no ice formation was observed in runs of about one  microsecond in previous work\cite{bresme2014,shevchuk12}. 

Regarding the possible existence of a liquid-liquid critical point, a key question is to know if the liquid can be equilibrated before
it freezes\cite{N_1992_360_00324_nolotengo,liu12,limmer:134503,nature_valeria,nature_st2_14}. In the case of the TIP4P/2005 model this is equivalent studiying whether 6 microseconds are enough to obtain the properties
of metastable water at high supercoolings (i.e in the range of 50 to 65K at 1 bar below the melting temperature).
Obviously the 6 microseconds refers to the study of this work (i.e at p=1bar). Further work is needed to analyze how 
$\tau_x$ changes with pressure. 

   In fig.\ref{tau_plot} the free energy barrier and size of the critical cluster are shown for TIP4P/2005 as a function of the supercooling. 
   Under  the conditions where the crystallization time from Avrami's expression is at a minimum we estimate a free 
   energy barrier of 14 $k_BT$ and a size of the critical cluster of about 60 molecules.

\section{ Conclusions }
  In this work we have determined the temperature at which several clusters become critical for both TIP4P and 
mW water models. In our previous work \cite{jacs2013} we performed similar calculations for TIP4P/2005 and TIP4P/ICE. By assuming 
that CNT can be used to describe the critical cluster size, the value of 
the interfacial Ih-water free energy $\gamma$ was obtained. 
We performed runs of the time evolution of the cluster size with time from its critical value and at the temperature
at which it is critical to determine the attachment rate $f^{+}$. Finally the value of the 
nucleation rate was estimated as a function of the supercooling, by using CNT 
to estimate the free energy barrier, and the attachment rate to obtain the kinetic prefactor. 
 The main conclusions of our work are:

\begin{enumerate}
 \item{ $\gamma$ was found to decrease with temperature with a slope  
 (related to the excess interfacial entropy) of about -0.25 mN/(K.m) in reasonable agreement with the previous estimate of Reinhard and 
Doye for the TIP4P/2005 model (i.e -0.18 mN/(K.m)).  For the mW the temperature dependence was found to be weaker. }

\item{ By extrapolating to the melting temperature an estimate of the interfacial free energy for the planar interface was 
obtained for several water models. The values of $\gamma$ for the planar interface decrease in the order  TIP4P/ICE, mW, TIP4P/2005 and TIP4P.
The values obtained  of $\gamma$ for the planar interface are in reasonable agreement with the reported experimental values $25-35$ mN/m.}

\item{ The attachment rate can be estimated quite well  by using the diffusion coefficient, and assuming a typical
attachment length of about one molecular diameter (i.e 3.5 \AA).  For the mW model , the decrease of D with T is weak, 
certainly accelerating significantly the dynamics at very low temperatures.}

\item{ By fitting the diffusion coefficient to an Arrhenius expression 
and assuming a linear variation of $\gamma$ with temperature we have estimated J for a wide range of temperatures. 
For the mW the values obtained for  J are in reasonable agreement with previous estimates. The predictions of the TIP4P/2005 for J 
are consistent (taking into account the uncertainties) with the experimental values. The model predicts a homogeneous nucleation 
temperature of about 37K , in agreement with experiments. }
\item{ At $T_{H}^{exp}$ the kinetic prefactor to be used in CNT should be of the order 
of 10$^{37}$$(m^{-3}s^{-1})$ whereas the free energy barrier $\Delta G_c$ is of about 53 $k_BT$. At $T_{H}^{sim}$, 
$\Delta G_c$ is of about  14 $k_BT$.}

\item{ The growth of ice is not arrested at least for temperatures up to 50K below the melting point. 
By using Avrami's equation we estimated that for large systems (i.e large enough to have at least one critical cluster in the simulation box) 
about 6 microseconds would be required to have a significant fraction of ice for a supercooling of about 60K. 
For smaller systems the time would be larger as one needs to wait until a critical cluster is formed.
Thus, for small systems, the liquid phase gains kinetic stability so it becomes possible to have the liquid as metastable phase for longer times.}
\end{enumerate}

We recognize that the picture provided in this work is far from complete, since we are using a number of approximations 
in the entire formulation. However it provides an initial framework for forthcoming studies possibly using more 
sophisticated methods such as umbrella sampling, forward flux sampling or transition path sampling.\cite{JPCM_2000_12_0A147} These calculations 
will be of much  interest, but certainly not cheap from a computational point of view. 
Although nucleation studies of ice from simulation are still in its infancy we hope our work will encourage further interest in 
this area, highly relevant for cryopreservation\cite{cryopres}, the food industry\cite{maki74} and climate prediction\cite{review_ice_formation_clouds_2005,baker97,demott10}.

\section{Acknowledgements}

This work was funded by grants FIS2013/43209-P of the MEC, and by the Marie Curie Integration Grants 303941-ANISOKINEQ-FP7-PEOPLE-2011-CIG  and 322326-
COSAAC-FP7-PEOPLE-2012-CIG .  C.Valeriani acknowledges financial support from a Juan de La Cierva and E.S. from a Ramon y Cajal Fellowship, respectively.
Calculations  were carried out thank to the supercomputer facility  Tirant  from
the Spanish Supercomputing Network (RES) (through project QCM-2014-1-0021). The authors thank Dr. Philip Geiger and Prof. Christoph 
Dellago for having kindly shared with them  configurations of the TIP4P/ICE. We thank Prof. Molinero for providing us the input files 
to perform runs of the mW model of water using LAMMPS. We thank the two reviewers of this paper for their helpful comments.  
We thank Dr. Carl McBride for a critical reading of the manuscript. 

%

\clearpage
\appendix

\section{Critical cluster}

Here, we show the performed trajectories done for the TIP4P and mW water models to establish in what degree of supercooling the cluster inserted becomes critical.
The performed trajectories for the TIP4P/2005 and TIP4P/ICE water models for the same purpose are shown in the Supplementary material of Ref [E. Sanz, C. Vega, J. R. Espinosa, R. Caballero-Bernal, J. L. F. Abascal and C. Valeriani, JACS, \textbf{135}, 15008, (2013)].

\begin{table}[h!]
\caption{ $N_t$ is the total
number of molecules in the system (ice cluster + surrounding liquid water molecules)
and $N_c$ is the number of molecules of the inserted spherical ice cluster after equilibration of the interface.}
\label{systemsize}
\centerline{
\begin{tabular}{cccccc}
\hline
  $N_{t}$  & $N_{c}^{TIP4P/2005}$ &  $N_{c}^{TIP4P/Ice}$ & $N_{c}^{TIP4P}$ & $N_{c}^{mW}$  \\
\hline
 22712  &  600  & 600 & 600 & 600 \\
 76781  &  3170 & 3167 & 3170 & 3167 \\
 182585 &  7931 & 7926 & 7931 & 7926 \\
\hline
\end{tabular}}
\end{table}
\begin{figure}[h!]
\centering
\includegraphics[clip,scale=0.25]{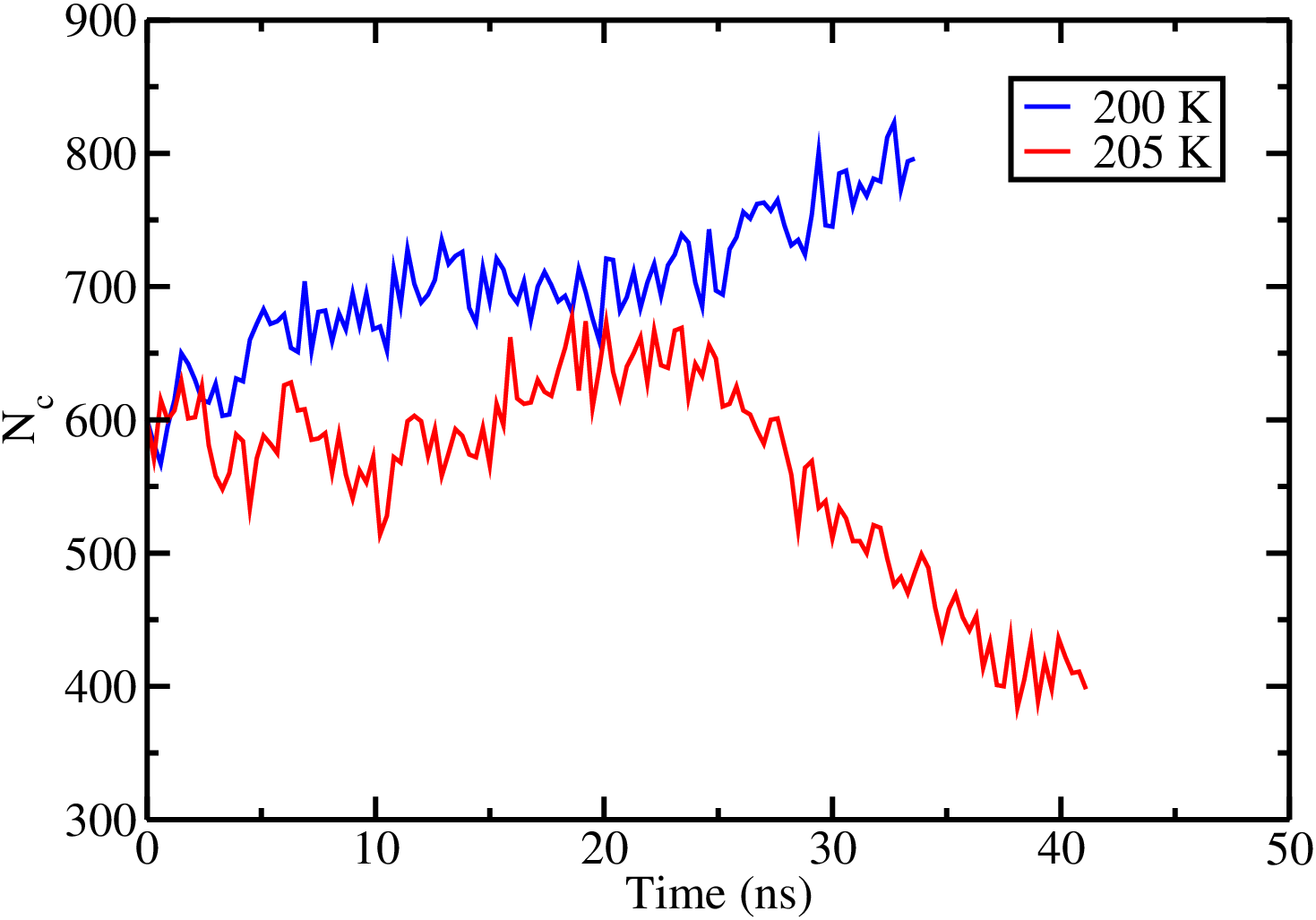}
\caption{Performed trajectories for the cluster of 600 ice molecules for the TIP4P model. The temperature for which this cluster is critical is 202.5 K.}
\label{600tc}
\end{figure}
\begin{figure}[h!]
\centering
\includegraphics[clip,scale=0.25]{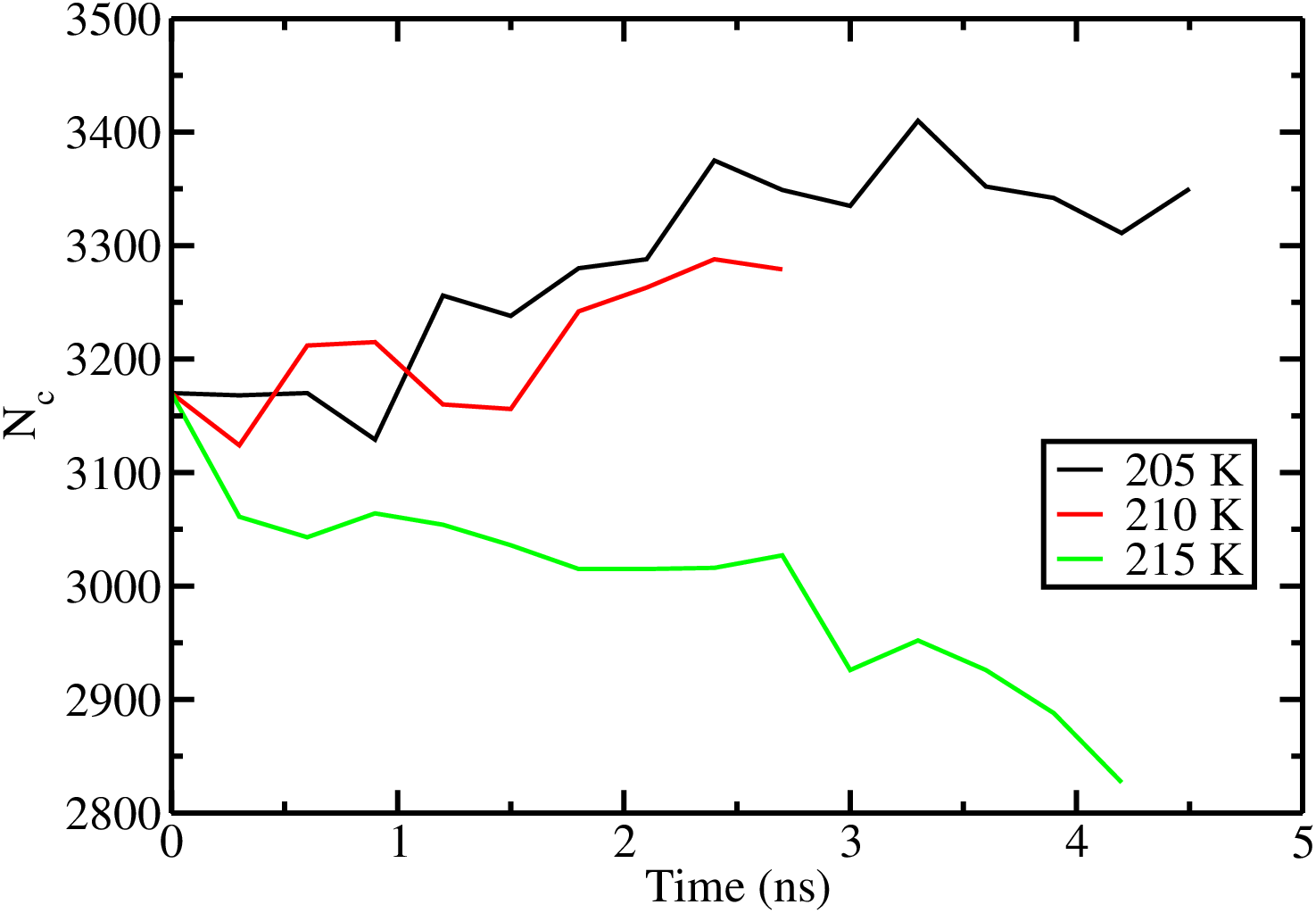}
\caption{Performed trajectories for the cluster of 3170 ice molecules for the TIP4P model. For this cluster size, the critical temperature is 212.5 K}
\label{3170tc}
\end{figure}
\begin{figure}[h!]
\centering
\includegraphics[clip,scale=0.25]{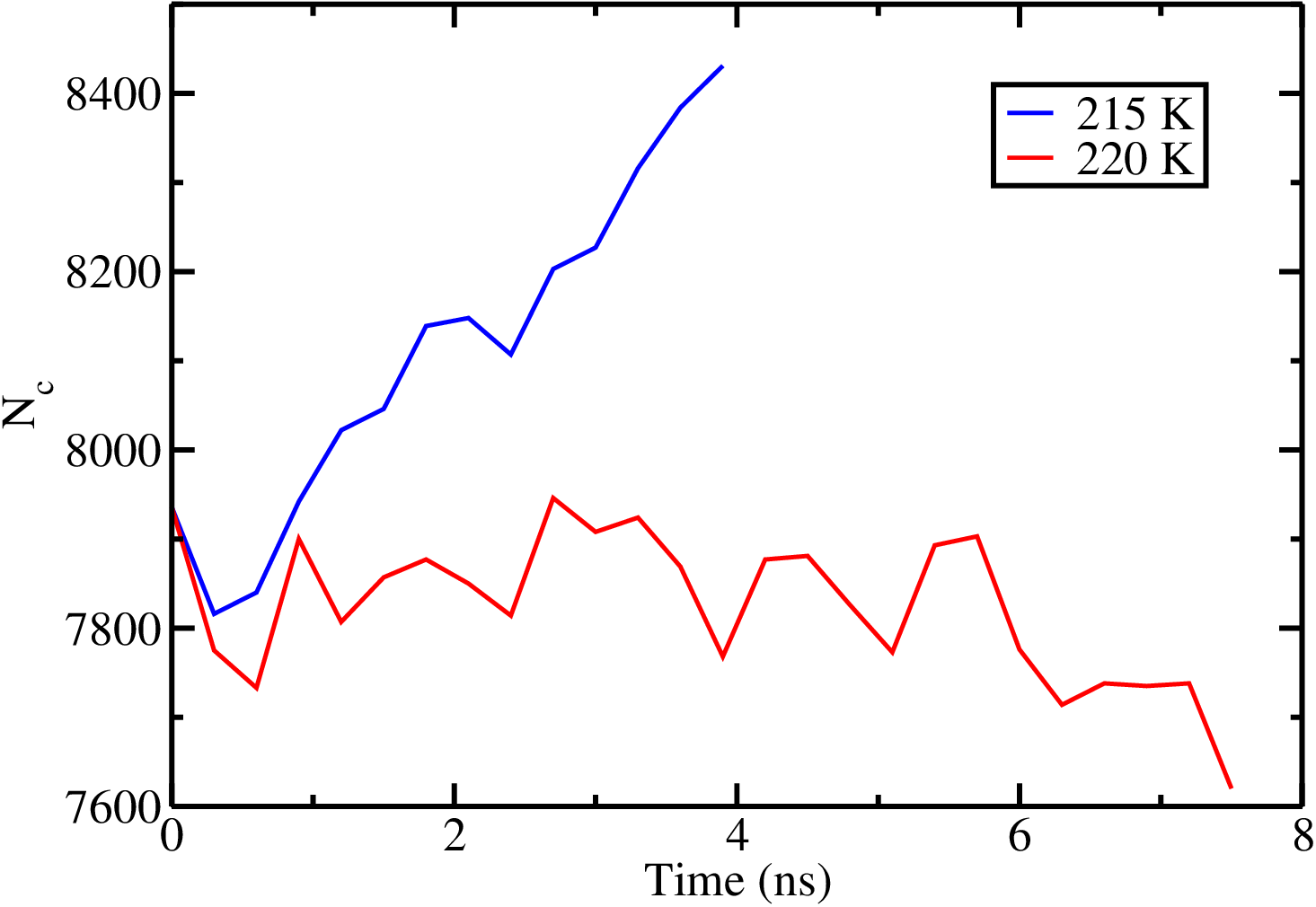}
\caption{Performed trajectories for the cluster of 7931 ice molecules for the TIP4P model. In this case the critical temperature is 217.5 K.}
\label{7931tc}
\end{figure}
\begin{figure}[h!]
\centering
\includegraphics[clip,scale=0.25]{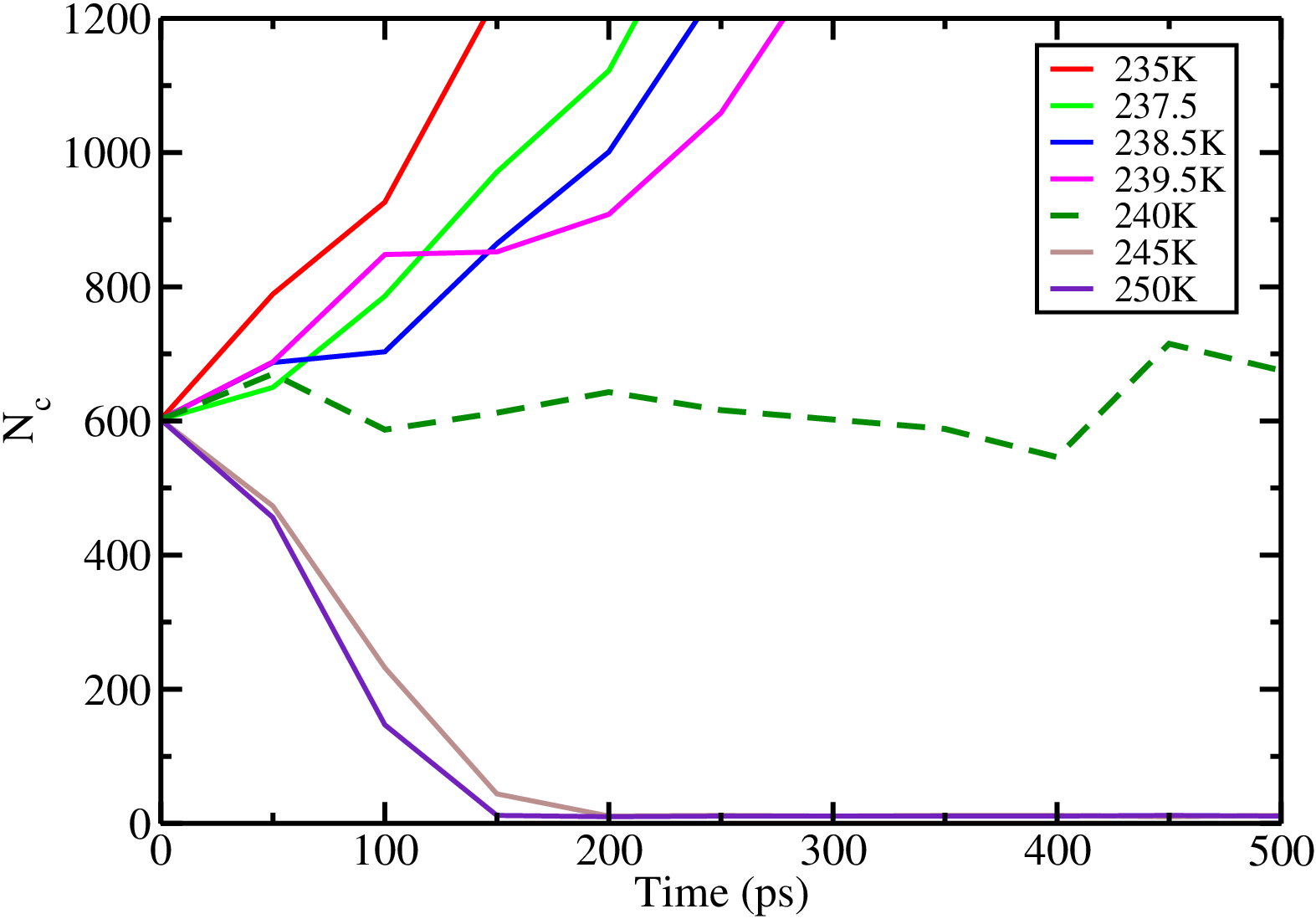}
\caption{Performed trajectories for the cluster of 600 ice molecules for the mW model. The temperature for which this cluster is critical is 240 K. }
\label{7931tc}
\end{figure}
\begin{figure}[h!]
\centering
\includegraphics[clip,scale=0.25]{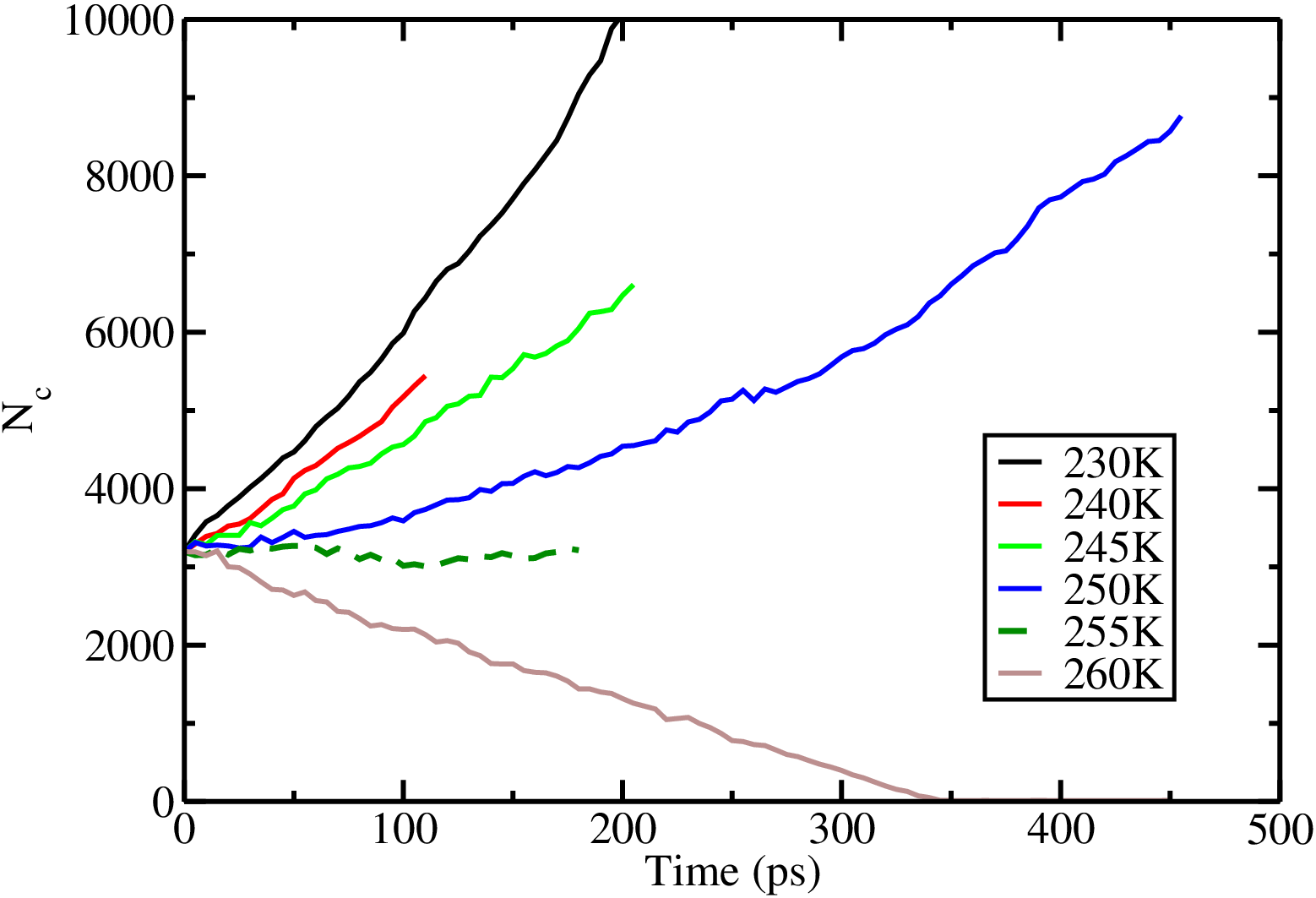}
\caption{Performed trajectories for the cluster of 3167 ice molecules for the mW model whose critical temperature is 255 K.}
\label{7931tc}
\end{figure}
\begin{figure}[h!]
\centering
\includegraphics[clip,scale=0.25]{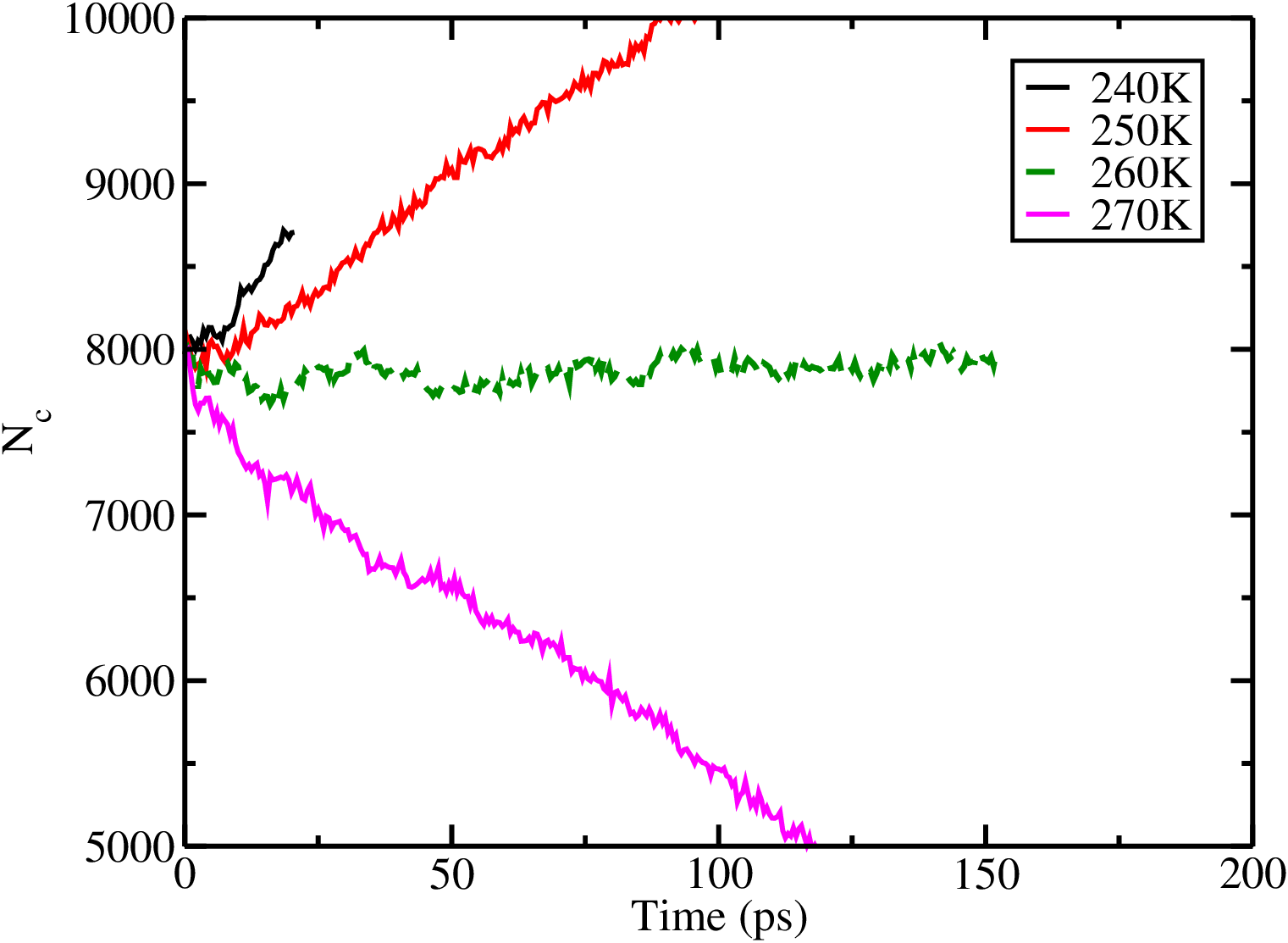}
\caption{Performed trajectories for the cluster of 7926 ice molecules for the mW model. The temperature for which this cluster is critical is 260 K.}
\label{7931tc}
\end{figure}

\clearpage

\section{Attachment rates}

\begin{figure}[h!]
\centering
\includegraphics[clip,scale=0.3]{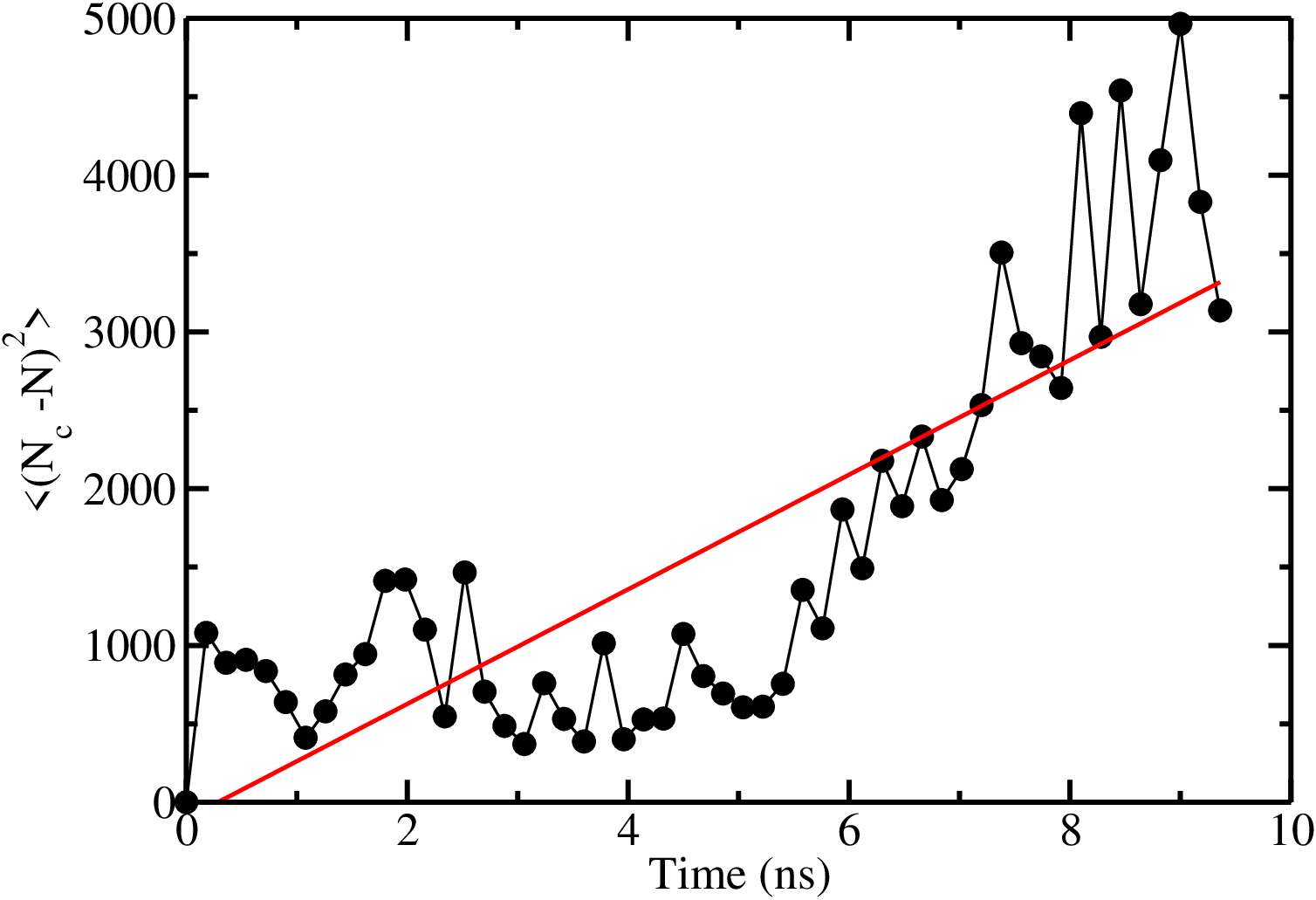}
\caption{Attachment rate calculated for the cluster of 600 ice molecules for the TIP4P model. 10 trajectories were performed under conditions of T=202.5 K and P=1bar.  }
\label{7931tc}
\end{figure}

\begin{figure}[h!]
\centering
\includegraphics[clip,scale=0.3]{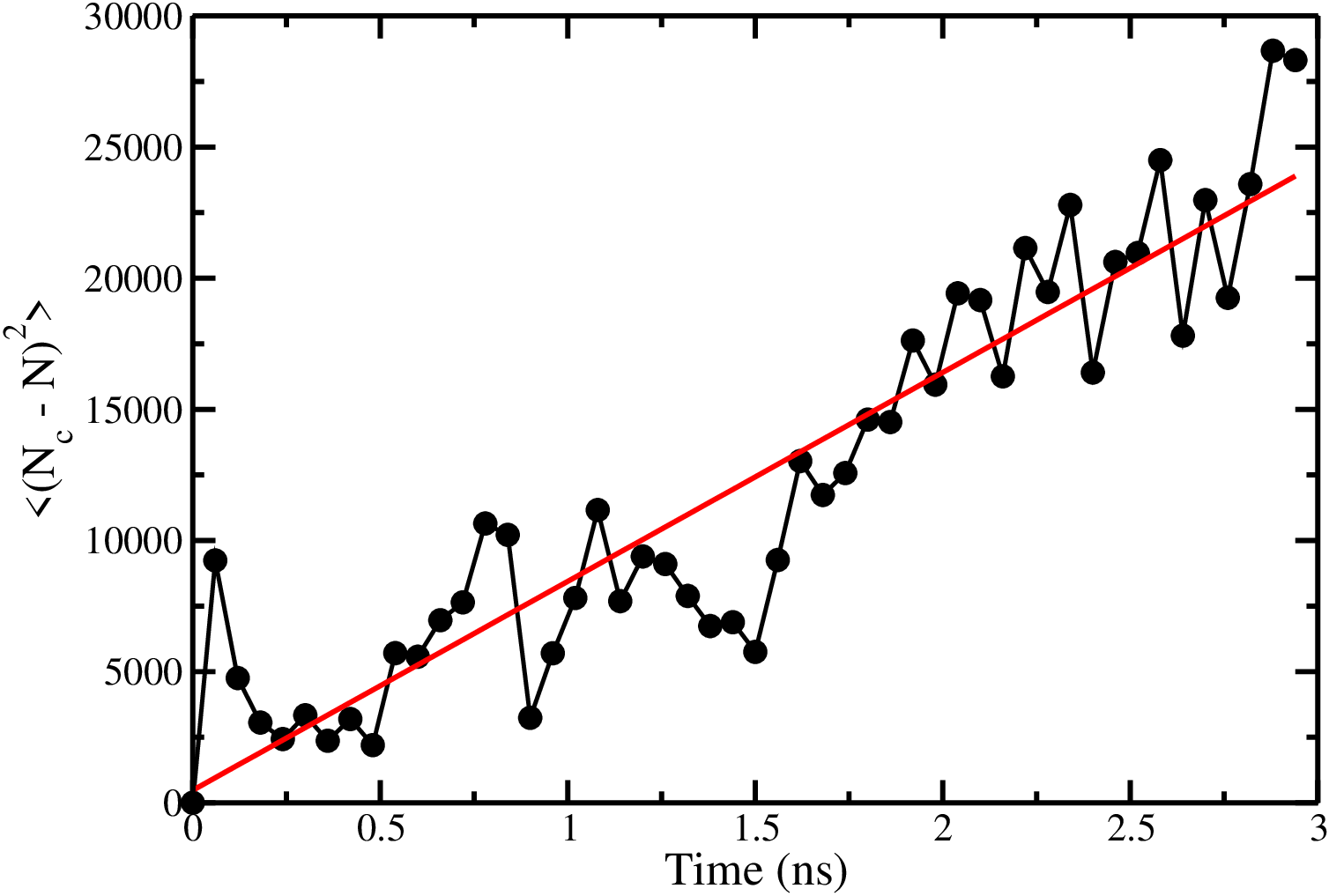}
\caption{Attachment rate calculated for the cluster of 3170 ice molecules for the TIP4P model. 10 trajectories were performed under conditions of T=212.5 K and P=1bar.}
\label{7931tc}
\end{figure}

\begin{figure}[h!]
\centering
\includegraphics[clip,scale=0.3]{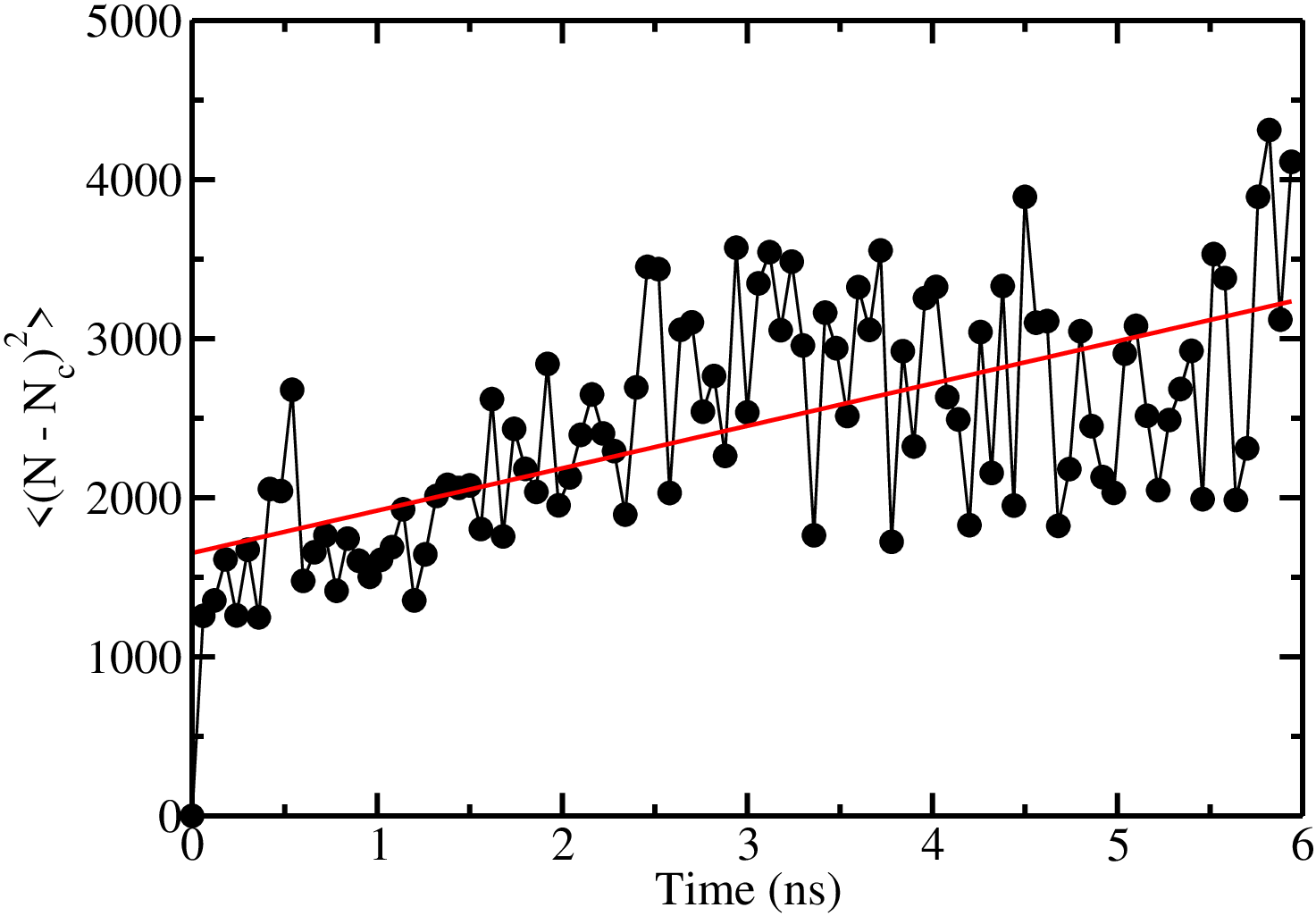}
\caption{Attachment rate calculated for the cluster of 600 ice molecules for the TIP4P/2005 model. 10 trajectories were performed under conditions of T=222.5 K and P=1bar. }
\label{7931tc}
\end{figure}

\begin{figure}[h!]
\centering
\includegraphics[clip,scale=0.3]{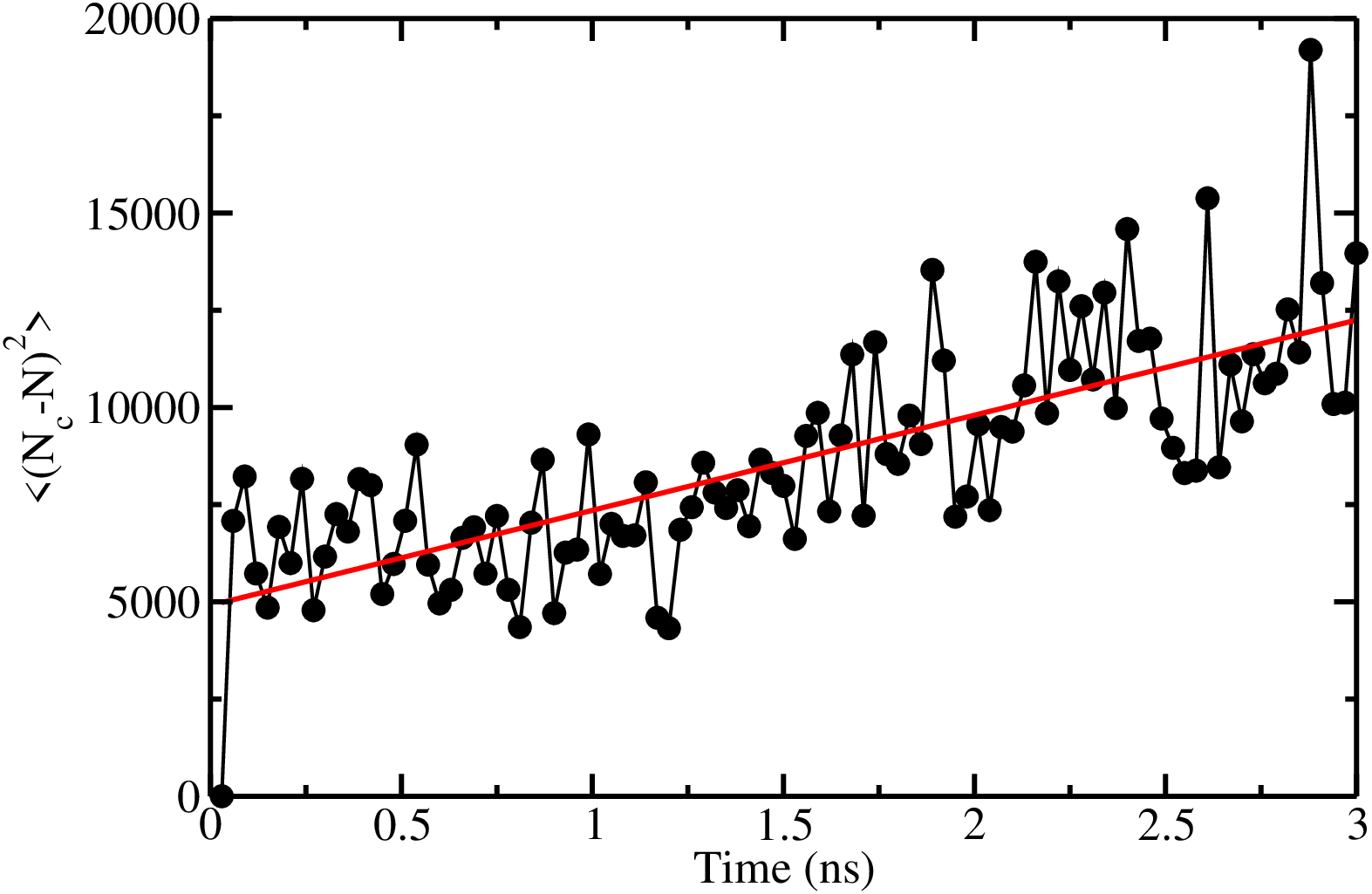}
\caption{Attachment rate calculated for the cluster of 3170 ice molecules for the TIP4P/2005 model. 10 trajectories were performed under conditions of T=232.5 K and P=1bar.}
\label{7931tc}
\end{figure}

\begin{figure}[h!]
\centering
\includegraphics[clip,scale=0.3]{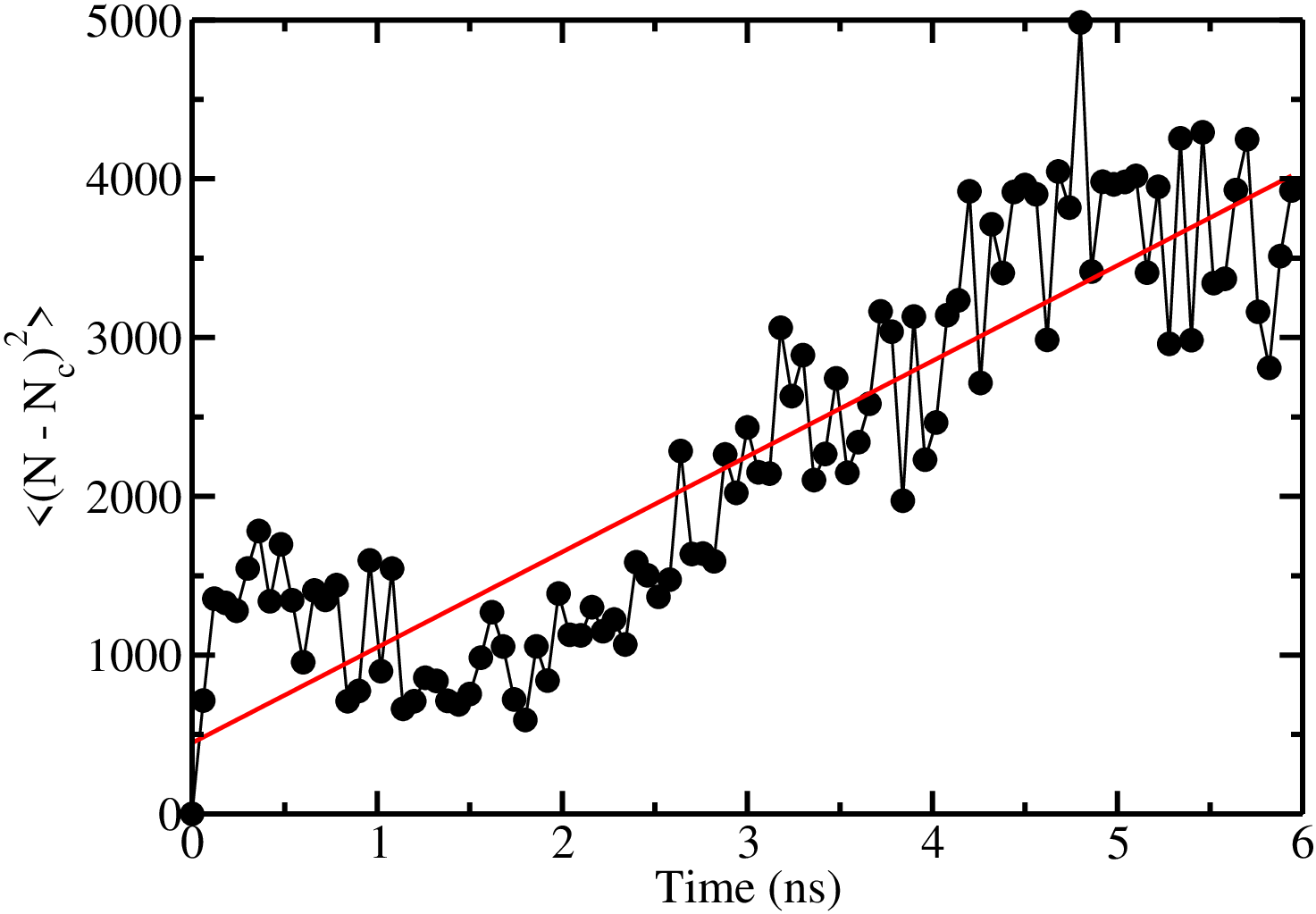}
\caption{Attachment rate calculated for the cluster of 600 ice molecules for the TIP4P/ICE model. 10 trajectories were performed under conditions of T=237.5 K and P=1bar.}
\label{7931tc}
\end{figure}

\begin{figure}[h!]
\centering
\includegraphics[clip,scale=0.3]{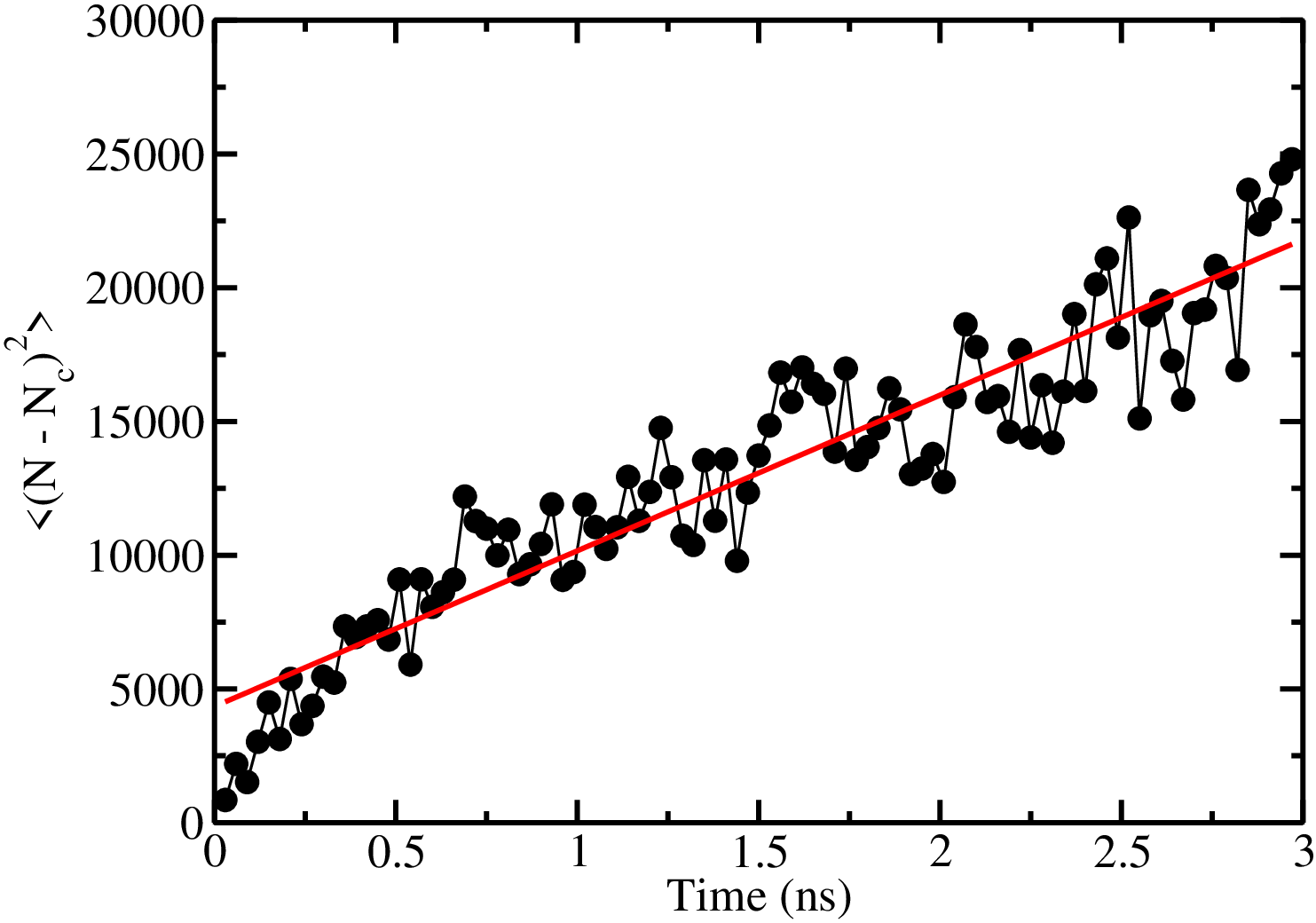}
\caption{Attachment rate calculated for the cluster of 3167 ice molecules for the TIP4P/ICE model. 10 trajectories were performed under conditions of T=252.5 K and P=1bar.}
\label{7931tc}
\end{figure}

\begin{figure}[h!]
\centering
\includegraphics[clip,scale=0.3]{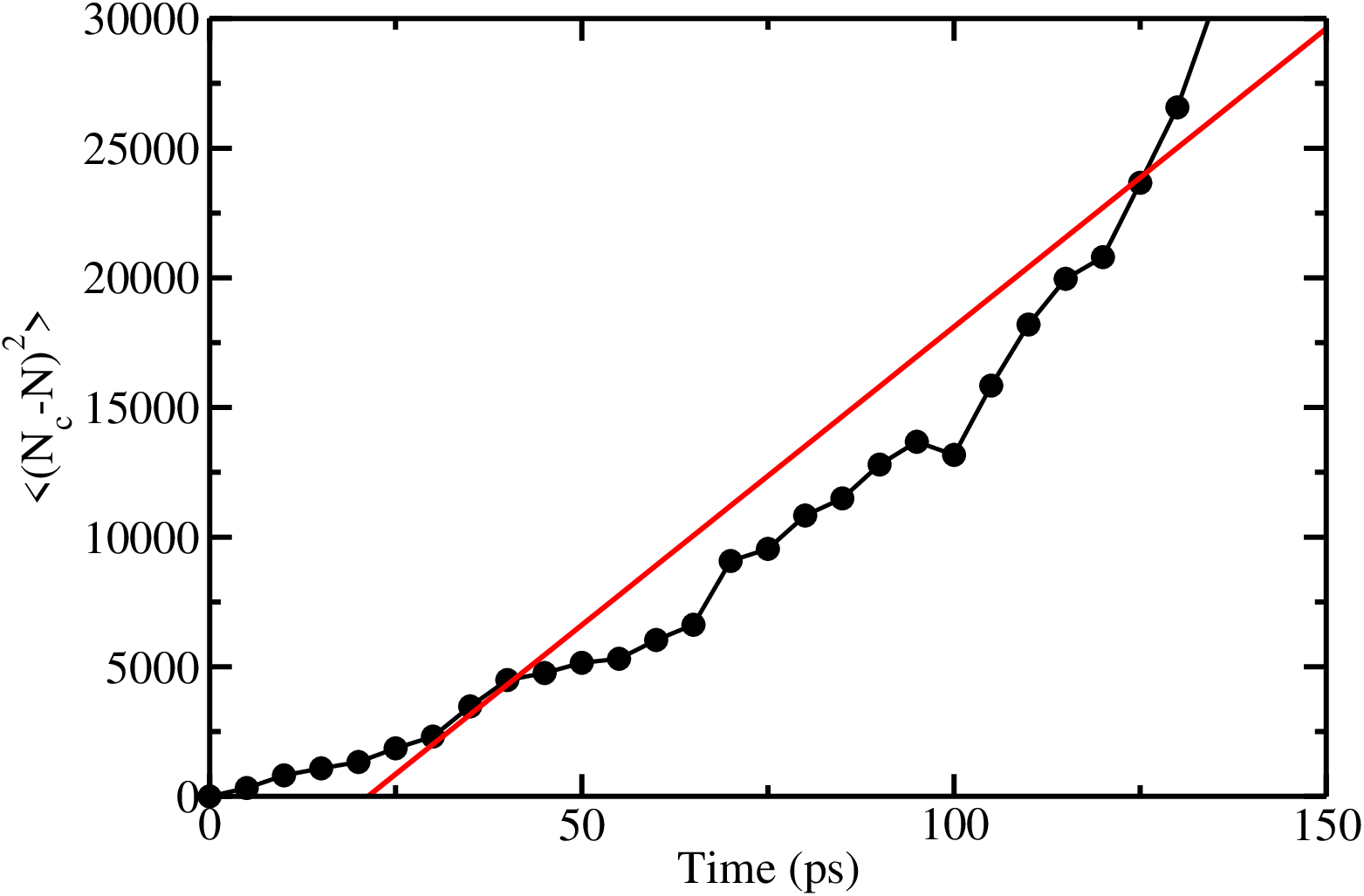}
\caption{Attachment rate calculated for the cluster of 600 ice molecules for the mW model. 30 trajectories were performed under conditions of T=240 K and P=1bar.}
\label{7931tc}
\end{figure}

\begin{figure}[h!]
\centering
\includegraphics[clip,scale=0.3]{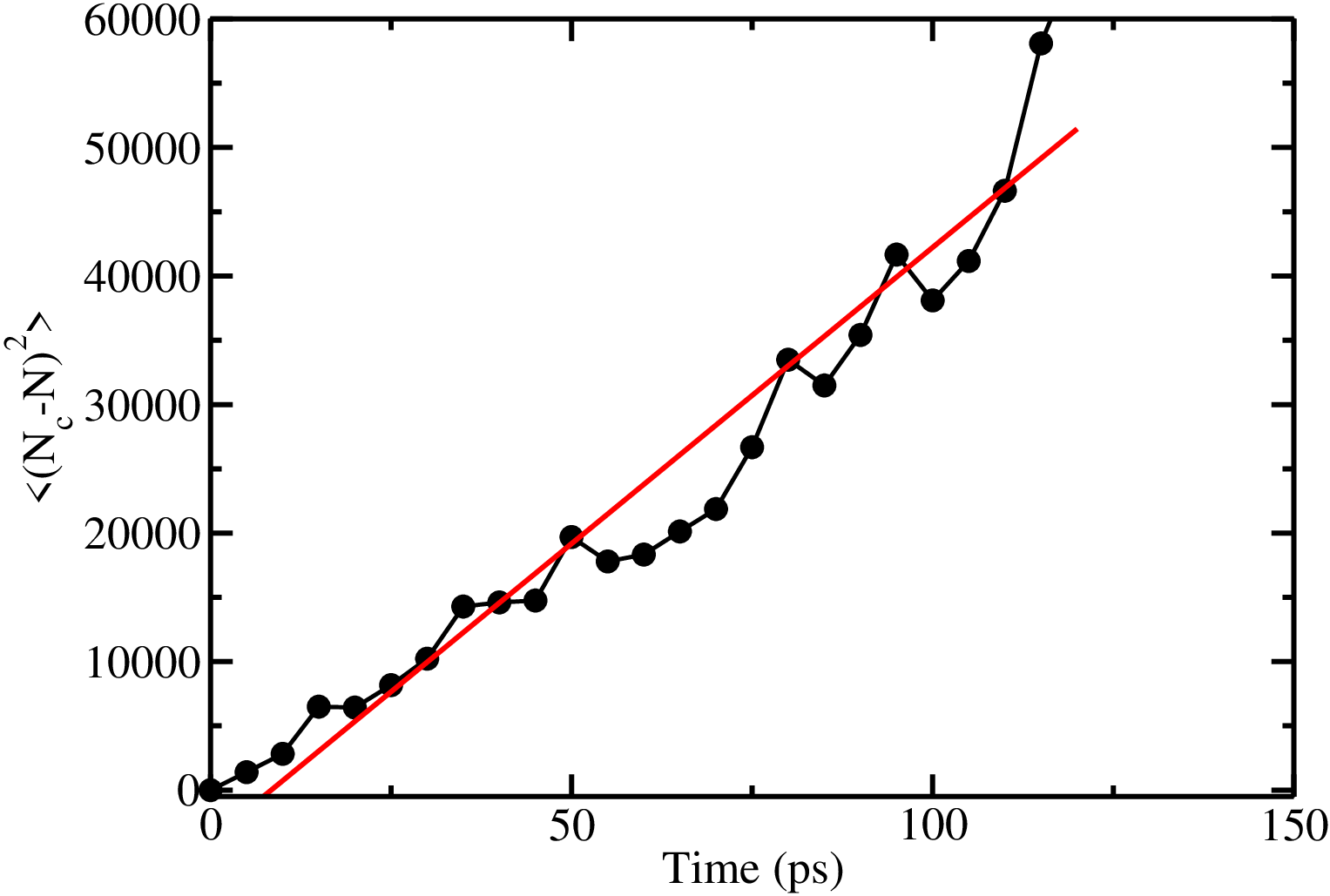}
\caption{Attachment rate calculated for the cluster of 3167 ice molecules for the mW model. 30 trajectories were performed under conditions of T=255 K and P=1bar.}
\label{7931tc}
\end{figure}

\begin{figure}[h!]
\centering
\includegraphics[clip,scale=0.3]{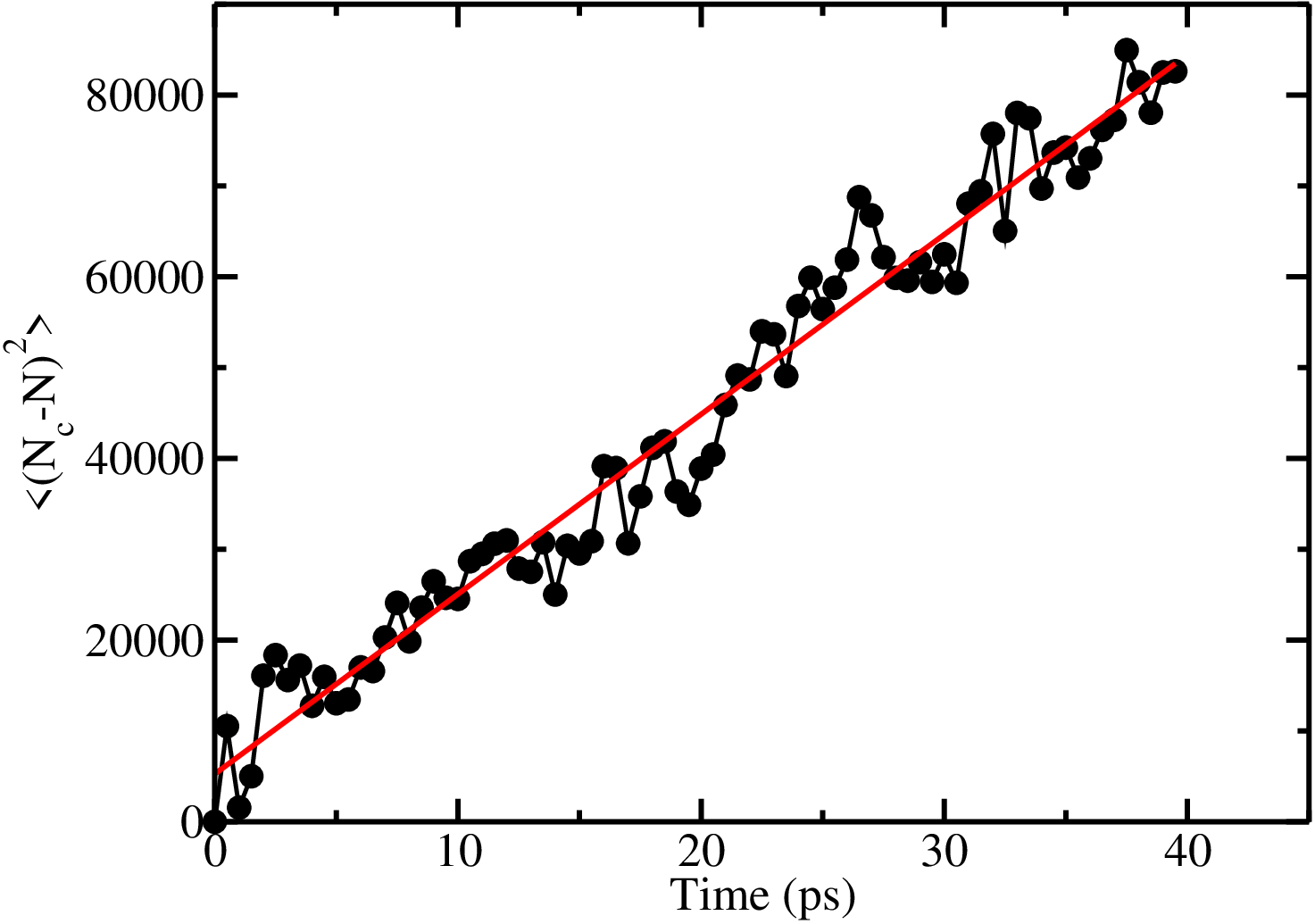}
\caption{Attachment rate calculated for the cluster of 7926 ice molecules for the mW model. 30 trajectories were performed under conditions of T=260 K and P=1bar.}
\label{7931tc}
\end{figure}

\end{document}